\documentclass{mn2ex} 
\usepackage{psfig}

\title[2dFGRS: Spectra and redshifts] 
{The 2dF Galaxy Redshift Survey: Spectra and redshifts}

\author[Colless et al.]{
\parbox[t]{\textwidth}{
Matthew Colless$^1$, 
Gavin Dalton$^2$,
Steve Maddox$^3$,
Will Sutherland$^4$,
Peder Norberg$^5$,
Shaun Cole$^5$,
Joss Bland-Hawthorn$^6$,
Terry Bridges$^6$, 
Russell Cannon$^6$, 
Chris Collins$^7$, 
Warrick Couch$^8$, 
Nicholas Cross$^4$,
Kathryn Deeley$^8$, 
Roberto De Propris$^8$,
Simon P.\ Driver$^4$, 
George Efstathiou$^9$, 
Richard S.\ Ellis$^{10}$, 
Carlos S.\ Frenk$^5$, 
Karl Glazebrook$^{11}$, 
Carole Jackson$^1$,
Ofer Lahav$^9$, 
Ian Lewis$^6$, 
Stuart Lumsden$^{12}$, 
Darren Madgwick$^9$,
John A.\ Peacock$^{13}$,
Bruce A.\ Peterson$^1$, 
Ian Price$^1$,
Mark Seaborne$^2$,
Keith Taylor$^{6,10}$ 
(the 2dFGRS team)}
\vspace*{6pt} \\ 
$^1$Research School of Astronomy \& Astrophysics, The Australian 
    National University, Weston Creek, ACT 2611, Australia \\
$^2$Department of Physics, University of Oxford, Keble Road, 
    Oxford OX1 3RH, UK \\
$^3$School of Physics \& Astronomy, University of Nottingham,
       Nottingham NG7 2RD, UK \\
$^4$School of Physics and Astronomy, University of St Andrews, 
    North Haugh, St Andrews, Fife, KY6 9SS, UK \\
$^5$Department of Physics, University of Durham, South Road, 
    Durham DH1 3LE, UK \\ 
$^6$Anglo-Australian Observatory, P.O.\ Box 296, Epping, NSW 2121,
    Australia\\  
$^7$Astrophysics Research Institute, Liverpool John Moores University,  
    Twelve Quays House, Birkenhead, L14 1LD, UK \\
$^8$Department of Astrophysics, University of New South Wales, Sydney, 
    NSW 2052, Australia \\
$^9$Institute of Astronomy, University of Cambridge, Madingley Road,
    Cambridge CB3 0HA, UK \\
$^{10}$Department of Astronomy, California Institute of Technology, 
    Pasadena, CA 91125, USA \\
$^{11}$Department of Physics \& Astronomy, Johns Hopkins University,
       Baltimore, MD 21218-2686, USA \\
$^{12}$Department of Physics, University of Leeds, Woodhouse Lane,
       Leeds, LS2 9JT, UK \\
$^{13}$Institute for Astronomy, University of Edinburgh, Royal Observatory, 
       Blackford Hill, Edinburgh EH9 3HJ, UK \\
}

\date{Accepted ---. Received ---; in original form ---.}

\newlength{\plotwidth}
\newlength{\fullwidth}
\newcommand{\tdfdr}{\mbox{\tt 2dfdr}}
\newcommand{\etal}{\mbox{et~al.}}
\newcommand{\bj}{\mbox{b$_{\rm\scriptscriptstyle J}$}}

\newcommand{\totphot}{467\,214}
\newcommand{\sgpspec}{193\,550}
\newcommand{\ngpspec}{139\,144}
\newcommand{\ranspec}{57\,019}
\newcommand{\totspec}{389\,713}

\newcommand{\h}{\mbox{$^{\rm h}$}}
\newcommand{\m}{\mbox{$^{\rm m}$}}

\newcommand{\persqdeg}{\mbox{\,deg$^{-2}$}}
\newcommand{\perpix}{\mbox{\,pixel$^{-1}$}}
\newcommand{\Mpc}{\mbox{$\,h^{-1}\,{\rm Mpc}$}}
\newcommand{\cubicMpc}{\mbox{$\,h^{-3}\,{\rm Mpc}^3$}}
\newcommand{\invMpc}{\mbox{$\,h\,{\rm Mpc}^{-1}$}}
\newcommand{\invcubicMpc}{\mbox{$\,h^3\,{\rm Mpc}^{-3}$}}
\newcommand{\kms}{\mbox{\,km\,s$^{-1}$}}
\newcommand{\CF}{\mbox{$\rm c_{\rm F}$}}

\newcommand{\xx}{\scriptsize \tt}
\newcommand{\plotone}[1]
           {\centering \leavevmode \psfig{file=#1,width=\plotwidth,clip=}}
\newcommand{\plottwo}[2]
           {\centering \leavevmode \psfig{file=#1,width=\plotwidth,clip=}
            \hfill \psfig{file=#2,width=\plotwidth,clip=}}
\newcommand{\plotfull}[2]
           {\centering \leavevmode \psfig{file=#1,width=#2\fullwidth,clip=}}
\newcommand{\plotrot}[3]
           {\centering \leavevmode \psfig{file=#1,width=#2\fullwidth,angle=#3,clip=}}

\newenvironment{mSQL}{\\ \tt }{\\}
 
\begin{document}

\maketitle

\begin{abstract}
The 2dF Galaxy Redshift Survey (2dFGRS) is designed to measure redshifts
for approximately 250\,000 galaxies. This paper describes the survey
design, the spectroscopic observations, the redshift measurements and
the survey database. The 2dFGRS uses the 2dF multi-fibre spectrograph on
the Anglo-Australian Telescope, which is capable of observing 400
objects simultaneously over a 2\degr\ diameter field. The source
catalogue for the survey is a revised and extended version of the APM
galaxy catalogue, and the targets are galaxies with extinction-corrected
magnitudes brighter than \bj=19.45. The main survey regions are two
declination strips, one in the southern Galactic hemisphere spanning
80$\degr$$\times$15$\degr$ around the SGP, and the other in the northern
Galactic hemisphere spanning 75$\degr$$\times$10$\degr$ along the
celestial equator; in addition, there are 99 fields spread over the
southern Galactic cap. The survey covers 2000\,deg$^2$ and has a median
depth of $\bar{z}$=0.11. Adaptive tiling is used to give a highly
uniform sampling rate of 93\% over the whole survey region. Redshifts
are measured from spectra covering 3600--8000\AA\ at a two-pixel
resolution of 9.0\AA\ and a median $S/N$ of 13\perpix. All redshift
identifications are visually checked and assigned a quality parameter Q
in the range 1--5; Q$\ge$3 redshifts are 98.4\% reliable and have an rms
uncertainty of 85\kms. The overall redshift completeness for Q$\ge$3
redshifts is 91.8\%, but this varies with magnitude from 99\% for the
brightest galaxies to 90\% for objects at the survey limit. The 2dFGRS
database is available on the WWW at http://www.mso.anu.edu.au/2dFGRS.
\end{abstract}


~ \\ 

\section{INTRODUCTION}
\label{sec:introduction}

The 2dF Galaxy Redshift Survey is designed to measure redshifts for
approximately 250\,000 galaxies in order to achieve an order of
magnitude improvement on previous redshift surveys, and provide a
detailed and representative picture of the galaxy population and its
large-scale structure in the nearby universe. The main goals of the
2dFGRS are:

1.~To measure the galaxy power spectrum $P(k)$ on scales up to a few
hundred Mpc, filling the gap between the small scales where $P(k)$ is
known from previous galaxy redshift surveys and the largest scales where
$P(k)$ is well-determined by observations of the cosmic microwave
background (CMB) anisotropies. Particular goals are to determine the
scale of the turnover in the power spectrum and to observe in the galaxy
distribution the acoustic peaks detected in the CMB power spectrum
(Percival \etal\ 2001).

2.~To measure the redshift-space distortion of the large-scale
clustering that results from the peculiar velocity field produced by the
mass distribution (Peacock \etal\ 2001). This distortion depends on both
the mass density parameter $\Omega$ and the bias factor $b$ of the
galaxy distribution with respect to the mass distribution, constraining
the combination $\beta=\Omega^{0.6}/b$.

3.~To measure higher-order clustering statistics of the galaxy
distribution in order to: (a)~determine the bias parameter $b$,
revealing the relationship between the distributions of mass and light
and yielding a direct measure of $\Omega$; (b)~establish whether the
galaxy distribution on large scales is a Gaussian random field, as
predicted by most inflationary models of the early universe; and
(c)~investigate the non-linear growth of clustering in the small-scale
galaxy distribution.

4.~To fully and precisely characterise the galaxy population in terms of
the distributions of fundamental properties such as luminosity, surface
brightness, spectral type and star-formation rate (Folkes \etal\ 1999;
Cole \etal\ 2001; Madgwick \etal\ 2001a; Cross \etal\ 2001).

5.~To quantify the relationships between the internal properties of
galaxies (such as luminosity, spectral type and star-formation rate) and
their external environment (the local density of galaxies and the
surrounding large-scale structure) in order to constrain models of
galaxy formation and evolution (Norberg \etal\ 2001a). 

6.~To investigate the properties of galaxy groups and clusters, not just
from existing cluster catalogues (De Propris \etal\ 2001), but by
defining a large, homogeneous sample of groups and clusters in redshift
space, avoiding the problems of cluster catalogues defined in projection
and allowing detailed study of the mass distributions and dynamical
evolution in the most massive bound structures in the universe.

7.~To provide a massive spectroscopic database for use in conjunction
with other surveys, for finding rare and interesting types of object,
and as a source for a wide variety of follow-up observations (Cole
\etal\ 2001; Sadler \etal\ 2001, Magliocchetti \etal\ 2001).

This paper provides an overview of the survey and detailed description
of the survey observations. The layout of the papers is as follows: \S2
summarises the main capabilities of the 2dF multi-fibre spectrograph;
\S3 describes the input source catalogue for the survey; \S4 discusses
the survey design, including the areas of sky covered by the survey and
the tiling of the survey fields; \S5 outlines the algorithm used to
assign fibres to targets, and its uniformity and completeness; \S6
describes the spectroscopic data obtained for the survey and the data
reduction methods; \S7 deals with the estimation of the redshifts and
various internal and external checks on their reliability and precision;
\S8 describes the survey masks, which encapsulate the coverage,
magnitude limits and redshift completeness of the survey; \S9 outlines
the main components and features of the survey database; \S10 summarises
some of the main results emerging from the survey; and \S11 provides
conclusions.

\section{THE 2dF SPECTROGRAPH}
\label{sec:instrument}

The survey is designed around the Two-degree Field (2dF) multi-fibre
spectrograph on the Anglo-Australian Telescope, which is capable of
observing up to 400 objects simultaneously over a 2\degr\ diameter field
of view. Here we summarise the aspects of the instrument relevant to the
2dFGRS; a full description is provided by Lewis \etal\ (2001) and the
2dF User Manual (http://www.aao.gov.au/2df). The four main components of
2dF are the corrector optics, the fibre positioner, the fibres and the
spectrographs.

The 2dF corrector optics are a four-element lens assembly which gives
1\arcsec\ images over a 2.1\degr\ diameter flat field of view at the
prime focus of the AAT. The corrector incorporates an atmospheric
dispersion compensator (ADC) which corrects for atmospheric dispersion
at zenith distances less than about 70\degr. The most significant
remaining image degradation is a chromatic variation in distortion,
which has the effect of dispersing images in the radial direction. This
effect is largest halfway out from the field centre, where the
wavelength range 0.35--1.0$\mu$m is dispersed over 2\arcsec\ radially.
The radial distortion introduced by the corrector gives an image scale
that varies over the field of view, from 15.5\arcsec/mm at the centre to
14.2\arcsec/mm at the edge; the corresponding change in focal ratio is
from f/3.4 to f/3.7.

The 2dF fibre positioner X-Y robot takes about 6--7~seconds on average
to position one fibre. Since approximately 550--580 fibre moves are
required to re-configure a typical 400-fibre field, this means that a
full re-configuration takes about 60--65~minutes. The actual
configuration time varies depending on the number of fibres used and the
complexity of the field. To avoid dead-time there are two field plates,
each with 400 fibres. While one field plate is placed at the focal plane
for observing, the other is being re-configured by the fibre positioner.
The positions of the field plates are reversed by tumbling them about
their horizontal axis.

The fibres are have 140$\mu$m diameter cores, corresponding to
2.16\arcsec\ at the field centre and 1.99\arcsec\ at the field edge.
There are 400 fibres on each field plate. The internal precision with
which the fibres are positioned is 11$\mu$m (0.16\arcsec) on average,
with no fibres outside 20$\mu$m (0.3\arcsec). The fibres terminate in
magnetic `buttons' that attach to the steel field plates. Each button
has a right-angle prism which directs the light from the focal plane
into the fibre. To prevent fibre buttons coming into direct contact and
protect the fragile prisms, the fibres cannot be placed too close
together. The absolute minimum fibre separation is 800$\mu$m
(approximately 12\arcsec), and the required separation is generally
larger (approximately 30\arcsec), as it depends on the detailed geometry
of the buttons and their relative orientations.

Half the fibres on each field plate go to one of the two identical
spectrographs. Each spectrograph has an f/3.15 off-axis Maksutov
collimator feeding a 150mm collimated beam to the grating and then, at
an angle of 40\degr, to an f/1.2 wide-field Schmidt camera. The
detectors are Tektronix CCDs with 1024$\times$1024 24$\mu$m pixels. The
200 fibre spectra imaged onto each CCD are separated by approximately 5
pixels. The 2dFGRS used the 300B gratings, which are blazed at 4200\AA\
and give a dispersion of 178.8\AA/mm (4.3\AA/pixel). For a typical
spectrograph focus of 2.1~pixels (FWHM), this corresponds to a FWHM
spectral resolution of 9.0\AA. The 2dFGRS observations used a central
wavelength around 5800\AA, and covered the approximate range
3600--8000\AA. Flexure is less than 0.2~pixel/hour (i.e.\ less than
40\kms\ over a typical integration time).

The overall system efficiency (source to detector) of 2dF with the 300B
gratings used in the 2dFGRS is 2.8\% at 4400\AA, 4.3\% at 5500\AA\ and
4.7\% at 7000\AA. These figures were obtained from measurements of
photometric standard stars, corrected to nominal 1\arcsec\ seeing.

\section{SOURCE CATALOGUE}
\label{sec:source}

The source catalogue for the survey (Maddox \etal\ 2001, in preparation)
is a revised and extended version of the APM galaxy catalogue (Maddox
\etal\ 1990a,b,c;1996). This catalogue is based on Automated Plate Measuring
machine (APM) scans of 390 plates from the UK Schmidt Telescope (UKST)
Southern Sky Survey. The extended version of the APM catalogue includes
over 5~million galaxies down to \bj=20.5 in both north and south
Galactic hemispheres over a region of almost 10$^4$\,deg$^2$ (bounded
approximately by declination $\delta$$\leq$+3\degr\ and Galactic
latitude $b$$\ge$30\degr). Small regions around bright stars, satellite
trails and plate flaws are excluded; these are accounted for by the
survey mask (see \S\ref{ssec:maglimmask}).

The \bj\ magnitude system for the Southern Sky Survey is defined by the
response of Kodak IIIaJ emulsion in combination with a GG395 filter. It
is zeropointed to Vega---i.e.\ \bj\ is equal to Johnson $B$ for an
object with zero colour in the Johnson--Cousins system. The colour
equation is normally taken to be
\begin{equation}
\bj = B - 0.28(B-V) ~,
\end{equation}
following Blair \& Gilmore (1982). A larger coefficient ($-0.35$) has
been suggested by Metcalfe \etal\ (1995), but we measure $-0.27 \pm
0.02$ in comparison with the ESO Imaging Survey (Arnouts \etal\ 2001),
and we therefore retain the usual value of $-0.28$. The photometry of
the catalogue is calibrated with numerous CCD sequences and, for
galaxies with \bj=17--19.45, has a 68\% spread of approximately
0.15\,mag, but with a non-Gaussian tail to the error distribution. We
emphasise that the calibration is to {\it total\/} CCD photometry, which
absorbs any remaining correction to the thresholded APM magnitudes.

The star-galaxy separation is as described in Maddox \etal\ (1990b),
and the locus dividing stars and galaxies was chosen to exclude as few
compact galaxies as possible while keeping the contamination of the
galaxy sample by stars to about 5\%. Spectroscopic identifications of
the survey objects (see \S\ref{sec:redshifts}), show that the stellar
contamination is in fact 6\%.

The source catalogue is incomplete at all magnitudes due to various
effects, including the explicit exclusion of objects classified by the
APM as merged images, the misclassification of some galaxies as stars,
and the non-detection (or misclassification as noise) of some low
surface brightness objects. This incompleteness has been studied in
comparisons with deeper wide-area CCD photometry by Pimbblet \etal\
(2001) and Cross \& Driver (2001, in preparation). The overall level of
incompleteness is 10--15\% and varies slightly with apparent magnitude,
being largest for the brightest and faintest objects. The main classes
of objects that are excluded are: (i)~merged galaxy images that are
explicitly excluded from the 2dFGRS source catalogue (about 60\% of the
missing objects); (ii)~large galaxies that are resolved into components
that are classified as stellar, merged or noise objects (20\%);
(iii)~compact normal galaxies that are detected but classified as stars
(15\%); and (iv)~low surface brightness galaxies that are either not
detected or classified as noise objects (5\%). Thus the main cause of
incompleteness is misclassification of objects rather than their
non-detection.

The target galaxies for the 2dFGRS were selected to have
extinction-corrected magnitudes brighter than \bj=19.45. Since the
targets were selected, improvements to the photometric calibrations and
revised extinction corrections have resulted in slight variations to the
magnitude limit over the survey regions---this is precisely quantified
by the magnitude limit mask for the survey (see
\S\ref{ssec:maglimmask}). The \bj\ extinction is taken to be
$A_{b_J}$==4.035$E$($B$$-$$V$), where the coefficient, and the reddening
$E$($B$$-$$V$) as a function of position, come from Schlegel \etal\
(1998). The limit of \bj=19.45 was chosen because: (i)~The surface
density of galaxies at \bj=19.45 (approximately 165\persqdeg) is
sufficiently larger than the surface density of 2dF fibres on the sky
(127\persqdeg) to allow efficient use of all fibres---few fibres are
unused even in low-density fields. (ii)~The time taken to configure a
typical field (60--65\,min) allows, with overheads, a sufficiently long
exposure time to reach the desired signal-to-noise level of
$S/N$$>$10\perpix\ for galaxies with \bj=19.45 even in rather poor
conditions. This limiting magnitude corresponds to a median redshift for
the survey of about $\bar{z}$=0.1, so that the 2dFGRS is essentially a
survey of the {\em local} universe.

\section{SURVEY DESIGN}
\label{sec:design}

\subsection{Survey areas}
\label{sec:areas}

\begin{figure*}
\plotfull{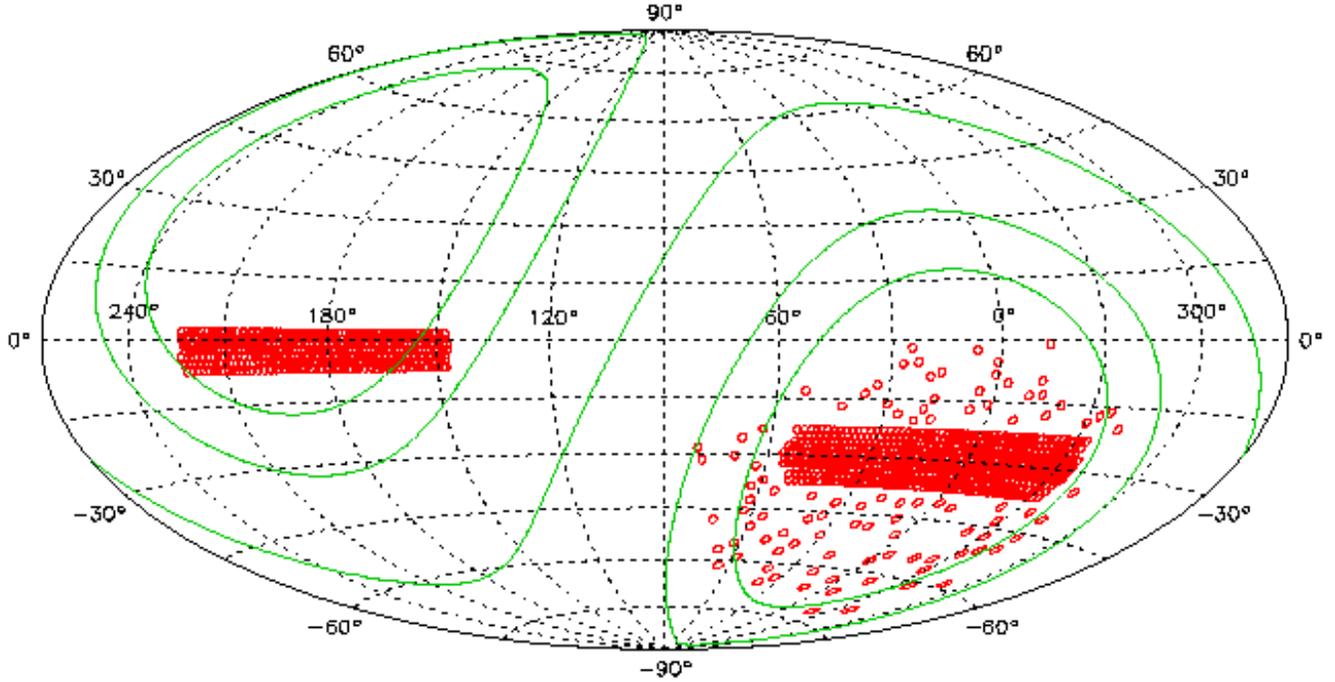}{1.0}
\caption{The 2dFGRS regions shown in an Aitoff projection of R.A.\ and
Dec., with individual 2dF fields marked as small circles. Also shown are
the lines of Galactic latitude $|b|$=0\degr, 30\degr, 45\degr. The
numbers of survey galaxies in these regions are: \sgpspec\ in the 643
fields of the 80$\degr$$\times$15$\degr$ SGP strip, \ngpspec\ in the 450
fields of the 75$\degr$$\times$10$\degr$ NGP strip, and \ranspec\ in the
99 fields scattered around the SGP strip.
\label{fig:skymap}}
\end{figure*}

\begin{figure*}
\plottwo{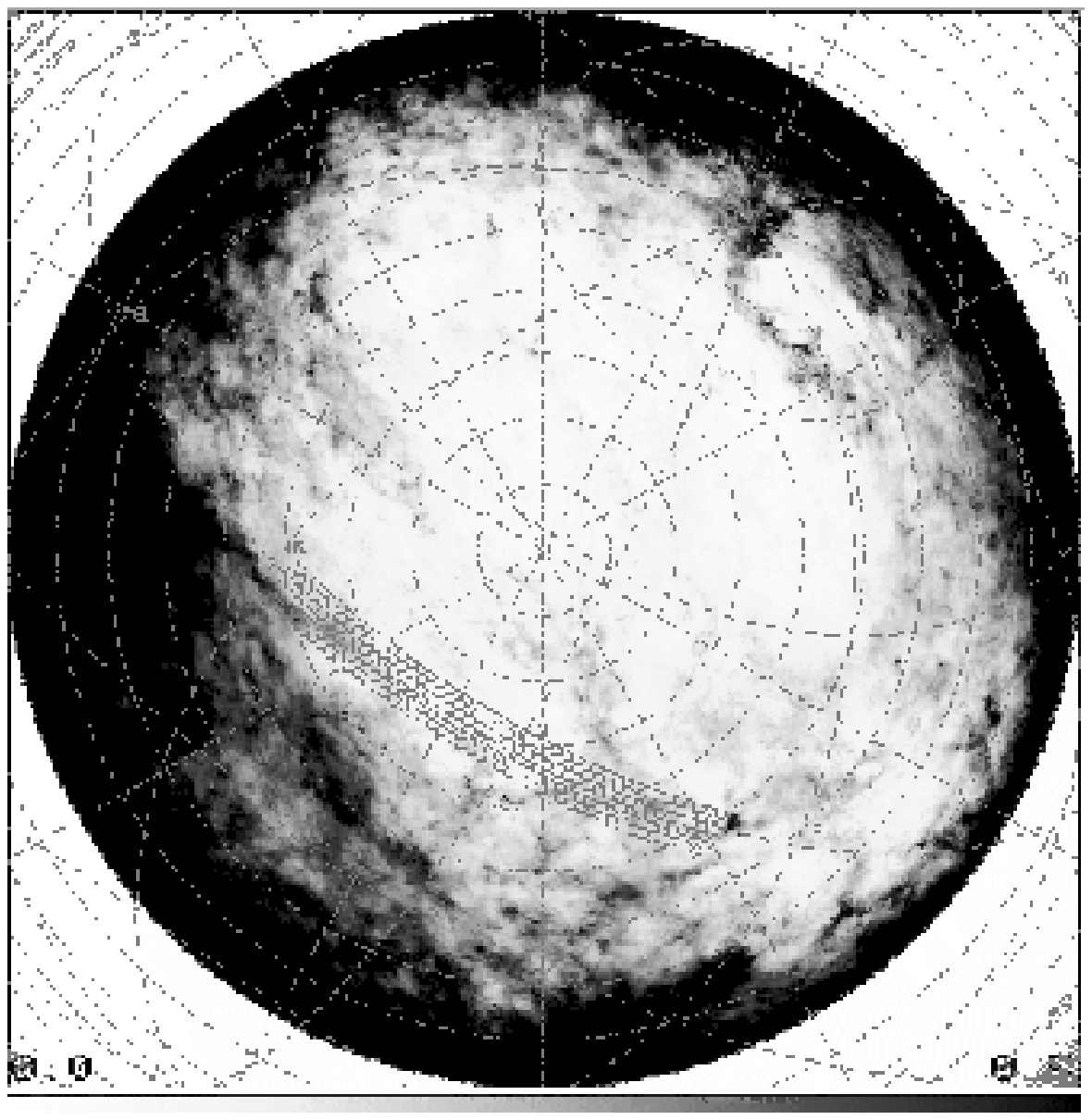}{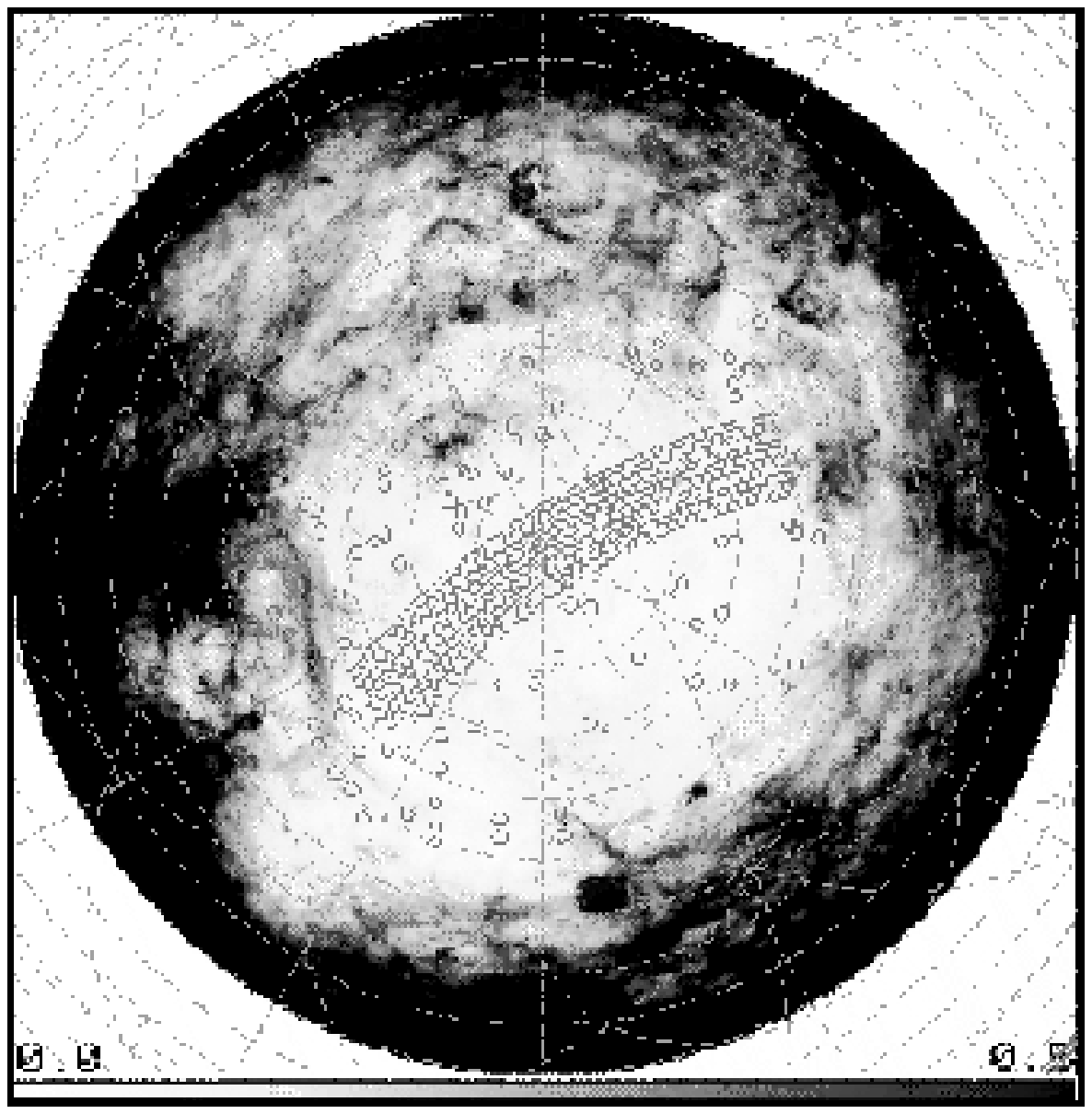}
\caption{The survey fields in the NGP (left) and SGP (right) on maps of
the extinction $A_{b_J}$ derived from Schlegel \etal\ (1998).
\label{fig:dustmap}}
\end{figure*}

The areas of the sky covered by the survey were chosen so as to satisfy
a number of different requirements. The first goal was to cover as large
a volume as possible, in order to closely approach a statistically
representative sample of the universe on the largest possible scales.
The second was to obtain near-complete sampling down to the survey limit
in order to have the finest possible resolution of structure on small
scales. The third requirement was to match the sample to the
observational capabilities of the 2dF instrument in order to achieve
high efficiency. The adopted geometry is an effective compromise between
these requirements.

The survey consists of two separate declination strips of overlapping
2\degr\ fields plus 99 scattered `random' 2\degr\ fields. One strip (the
SGP strip) is in the southern Galactic hemisphere and covers
approximately 80$\degr$$\times$15$\degr$ centred close to the South
Galactic Pole ($21\h40\m<\alpha<03\h40\m$,
$-37.5\degr<\delta<-22.5\degr$).
The other strip (the NGP strip) is in the northern Galactic hemisphere
and covers 75$\degr$$\times$10$\degr$ 
($09\h50\m<\alpha<14\h50\m$, $-7.5\degr<\delta<+2.5\degr$). 
The 99 `random' fields are chosen from the low-extinction region of the
APM catalogue in the southern Galactic hemisphere outside the survey
strip (the mean extinction over each field is required to be less than
0.2\,mag---see Figure~2 of Efstathiou \& Moody (2001)). The fields are
chosen pseudo-randomly within this region, except that the field centres
are at least 3\degr\ apart. A map of the survey fields on the sky is
shown in Figure~\ref{fig:skymap}; the locations of the fields with
respect to the extinction map derived from Schlegel \etal\ (1998) are
shown in Figure~\ref{fig:dustmap}. All the survey fields lie at Galactic
latitudes greater than $|b|$=30\degr, and the whole of the SGP strip and
most of the NGP strip and the random fields lie at Galactic latitudes
greater than $|b|$=45\degr.

\begin{figure}
\plotone{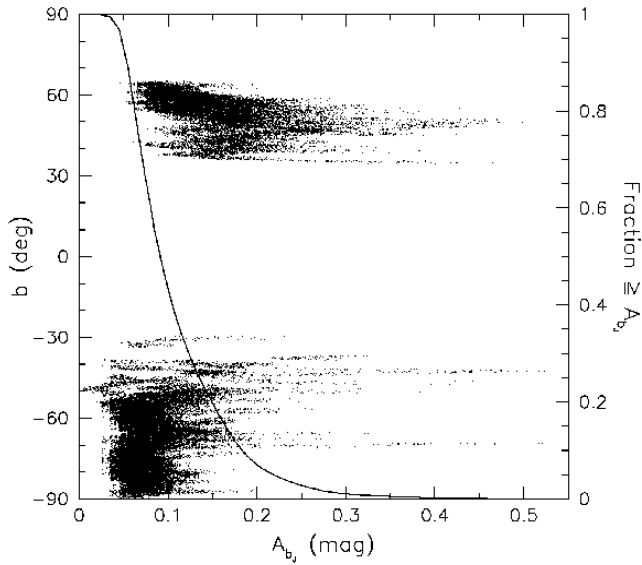}
\caption{The distribution of extinction corrections $A_{b_J}$ with
Galactic latitude $b$ (dots and left axis) and the fraction of
corrections larger than a given value (line and right axis).
\label{fig:extndist}}
\end{figure}

The distribution of extinction corrections as a function of Galactic
latitude, and the fraction of corrections larger than a given value, are
shown in Figure~\ref{fig:extndist}. Overall, the median correction is
0.07\,mag, 90\% are less than 0.16\,mag, and 99\% are less than
0.26\,mag; the corresponding quantiles in the NGP are
(0.12,0.19,0.28)\,mag, in the SGP (0.05,0.07,0.11)\,mag and in the
random fields (0.07,0.13,0.30)\,mag.

The 2dFGRS target sample of galaxies contains \sgpspec\ galaxies in the
SGP strip, \ngpspec\ galaxies in the NGP strip, and \ranspec\ galaxies
in the random fields. This gives a total of \totspec\ possible targets,
significantly more than the survey goal of 250\,000 galaxies. Survey
observations of the NGP and SGP strips are proceeding outwards in
declination from the centre of each strip towards this goal. Note that
the total number of galaxies listed in the survey source catalogue (and
the survey database) is \totphot, which is larger than the number of
possible survey targets because the source catalogues for the NGP and
SGP strips conservatively includes galaxies fainter than the
spectroscopic survey magnitude limit, down to \bj=19.6.

At the median redshift of the survey ($\bar{z}$=0.11) the SGP strip
extends over 400\Mpc$\times$75\Mpc, and the NGP strip over
375\Mpc$\times$50\Mpc. Out to the effective limit of the survey at
$z$$\approx$0.3, the strips contain a volume of
1.2$\times$10$^8$\cubicMpc (for $\Omega_m$=0.3, $\Omega_{\Lambda}$=0.7);
the volume sparsely sampled including the random fields is between two
and three times larger.

\subsection{Tiling the survey} 
\label{ssec:tiling} 

The survey limit of \bj=19.45 was chosen, in part, because it gives a
good match between the surface density of galaxies and the surface
density of 2dF fibres. Due to clustering, however, the number of
galaxies in a given field varies considerably. The rms variation in the
number of galaxies per randomly-placed 2\degr\ field is 140 at \bj=19.5,
and is largely independent of the choice of magnitude limit over the
range considered here. To make efficient use of 2dF we therefore require
an algorithm for tiling the sky with 2\degr\ fields that allows us to
cover the survey area at a high, and nearly uniform, sampling rate with
the minimum number of 2dF fields.

So long as the sampling of the source catalogue is not biased in any way
that depends on the photometric or spectroscopic properties of the
galaxies, we can always use the source catalogue to accurately determine
the sampling rate as a function of position (see \S\ref{sec:mask}). The
sampling can then be accounted for in any analysis. However to keep such
corrections to a minimum, considerable effort has been invested in
making the sampling as complete and uniform as possible.

There are a number of possible approaches to laying down target field
centres. The simplest is to adopt a uniform grid of equally spaced
centres and then either randomly sample each field with the number of
available fibres or observe each field several times until all the
galaxies have been observed. The second of these options is clearly
inefficient, as it will give rise to a large number of fields being
observed with significantly less than the full complement of fibres,
while the first is undesirable as it gives a different sampling factor
for each field. A more efficient solution is to use an adaptive tiling
strategy, where we allow each field centre to drift from the regular
grid so that we maximise the number of targets that are assigned to each
field, subject to the constraint that the number of targets assigned to
any one field should not exceed the number of available fibres, $N_f$.

We begin with a uniform grid with field centres equally spaced by
$s=\sqrt{3}\degr$ in right ascension along rows 1.5\degr\ apart in
declination. For each galaxy in the survey we then determine how many
fields it lies within, and assign each field a weight $w_i$, where
\begin{eqnarray}
w_i = & 0                   & {\rm if}~N_i \ge N_f \\
    = & 1 - \frac{N_i}{N_f} & {\rm if}~N_i < N_f \nonumber
\end{eqnarray}
and $N_i$ is the number of galaxies in this field. These weights are
normalised such that $\sum_1^{N_t}w_i = 1$, where $N_t$ is the number of
fields that this galaxy could be assigned to. The galaxy in question is
then randomly assigned to one of the fields using these weights, unless
all the fields are already filled, in which case it is assigned to the
first field in which it was found. Once all the field occupancies have
been determined in this way we move each field in right ascension by an
amount
\begin{eqnarray}
\delta\alpha_i = & \Delta\alpha+0.05s\frac{N_i}{N_f} & {\rm if}~(N_i \ge N_f)\\
= & \Delta\alpha-s^\prime\left(1-\frac{N_i}{N_f}\right) & {\rm if}~(N_i<N_f) \nonumber
\end{eqnarray}
where $s$ is the maximum allowed separation between adjacent fields,
$s^\prime$ is the current distance to the neighbouring field centre, and
$\Delta\alpha$ is the cumulative shift that has currently been applied
to this row. New fields are added at the end of each row if the total
length of the row has contracted enough to exclude any galaxies at the
trailing edge. We found that using a fixed separation of 1.5\degr\ in
declination, and adjusting the tile positions in right ascension only,
provided sufficient flexibility to achieve uniform high completeness
without a large increase in the total number of fields and without
leaving gaps in the sky coverage.

In practice, it is found that the above prescription requires a further
modification to account for the position of each object within the
field. This additional constraint arises because of the physical
restrictions on the positioning of individual fibres, both in terms of
their deviation from the radial angle and their extension from the
parked position at the edge of the field. We apply this constraint by
dividing each field into 36 sub-fields and restricting the number of
targets that can be assigned to each sub-field to 16. This is larger
than the number of fibres whose park positions fall within the arc of
the sector, but allows for the fact that the area of the field that can
be reached by each fibre is increasing with the fibre extension. Without
this extra constraint the algorithm tends to place large clusters in the
overlapping areas of neighbouring fields where, although there are more
available fibres because of the overlap, it rapidly becomes impossible
to use these fibres because of the high density of targets close to the
edge of each field. Limiting the number of targets within each sector
effectively removes this problem, and so increases the uniformity of the
survey completeness.

The adaptive tiling algorithm also needed to cope with the requirement
that the galaxy redshift survey be merged with the concurrent survey of
QSO candidates (the 2dF QSO Redshift Survey: Boyle \etal\ 2000; Croom
\etal\ 2001). This results in a higher surface density of targets in the
region of overlap, for which we compensate by reducing the separation in
declination of the tiling strips in the QSO survey regions to 75\% of
the original value (i.e.\ to 1.125\degr). As the QSO survey occupies the
central declination strip of our SGP survey region this results in a
3-4-3 arrangement of tiling strips over the three rows of UKST fields
used in the SGP survey. A similar consideration applied to the NGP
survey region gives a 3-4 arrangement if we consider the full areas of
the two rows of UKST fields used.

With these additional modifications, our tiling algorithm was able to
achieve an overall sampling rate of 99.8\%, dropping to 93\% after the
assignment of fibres to targets has been made (see
\S\ref{ssec:fibrecomplete}). On average, fewer than 5\% of the fibres
in each field were unused. For comparison, the mean sampling rate that
would be achieved for randomly distributed fields is 68\%, with an rms
variation of 15\%. 

\section{FIBRE ASSIGNMENT}
\label{sec:fibres}

The tiling procedure described above fixes the location of the 2dF
target fields on the sky and provides a first pass priority scheme
(based on the random field assignments) for use in assigning targets to
fibres. However, the mechanical constraints of 2dF imply that we cannot
usually allocate fibres to all the targets assigned to a particular
field, and so we consider all targets that lie within each field
boundary. Our tiling scheme implies that many of the survey targets are
found on more than one field, so we adopt a priority scheme as follows:
targets which are unique to the field in question are assigned the
highest priority, then targets which are assigned to the field, but
which can also be reached from neighbouring fields, then targets which
fall within the field, but which are assigned to a different,
overlapping field. The priorities of the QSO targets are increased one
step to ensure that we do not imprint the strong clustering pattern of
the galaxy distribution on the weak clustering expected from the QSO
sample.

\subsection{Single-field assignments}
\label{ssec:singlefield}

The allocation procedure for a single field can be divided into two
distinct steps: an initial allocation pass and a fibre swapping pass.
These two steps are repeated for each range of decreasing priorities for
targets that have been assigned to this field. An attempt is then made
to reduce the complexity of the configuration by identifying pairs of
allocated fibres which are crossed, and which can safely be exchanged.
The allocation and uncrossing steps are then repeated to allow
consideration of the lower priority targets which were not assigned to
this field. The details of these steps are as follows:

(1)~{\em Initial allocation pass.} For each target, $t$, of a given
priority we determine the set, ${\cal F}_t$, of all fibres that could be
allocated to that target in the absence of any other allocations. We
then determine the set, ${\cal T}_t$, of all targets which fall within
the sector defined by the target and the two most extreme fibres in
${\cal F}_t$. We then use the difference $N({\cal T}_t)-N({\cal F}_t)$
between the number of targets in ${\cal T}_t$ and the number of fibres
in ${\cal F}_t$ to determine which targets are hardest to access and
hence should be given the highest priority: targets within the sector
defined by the target with the largest value of $N({\cal T}_t)-N({\cal
F}_t)$ are allocated first. This procedure is then repeated for targets
of successively lower priorities.

(2)~{\em Fibre-swapping pass.} For each remaining unallocated target,
$u$, we now consider each fibre in ${\cal T}_u$. For each fibre in this
set that could still be allocated to $u$ given the current configuration
we make a tentative re-allocation of this fibre from its initial
assignment, $i$, to target $u$ and check to see if any unallocated
fibres can be assigned to $i$. If no unallocated fibres can be assigned
we repeat this process recursively until one of the following conditions
have been satisfied: either (i)~a previously unallocated fibre is
allocated (implying that $u$ has been allocated and that all previously
allocated targets remain allocated); or (ii)~the search exceeds a depth
of 10 iterations. In the latter case, if at any point in the search a
fibre has been moved from a low-priority target such that a target of
higher priority has been allocated then the search is unpacked to this
point; if no such trade-off was found then the search is unpacked to the
original configuration.

(3)~{\em Fibre-uncrossing pass.} Each pair of fibres in the
configuration is tested to see if the fibres cross. For each crossed
pair an attempt is made to swap the allocations of fibres to targets.
This process is iterated until no further swaps can be made. The net
effect of this procedure is that the final configuration supplied to the
positioner is simplified so that the transition to the {\it following}
configuration will require fewer fibres to be moved twice. This
simplification effectively reduces the time required for each
re-configuration, and is achieved without constraint on the allocation.

\subsection{Field-overlap assignments}
\label{ssec:fieldoverlap}

\begin{figure*}
\plotfull{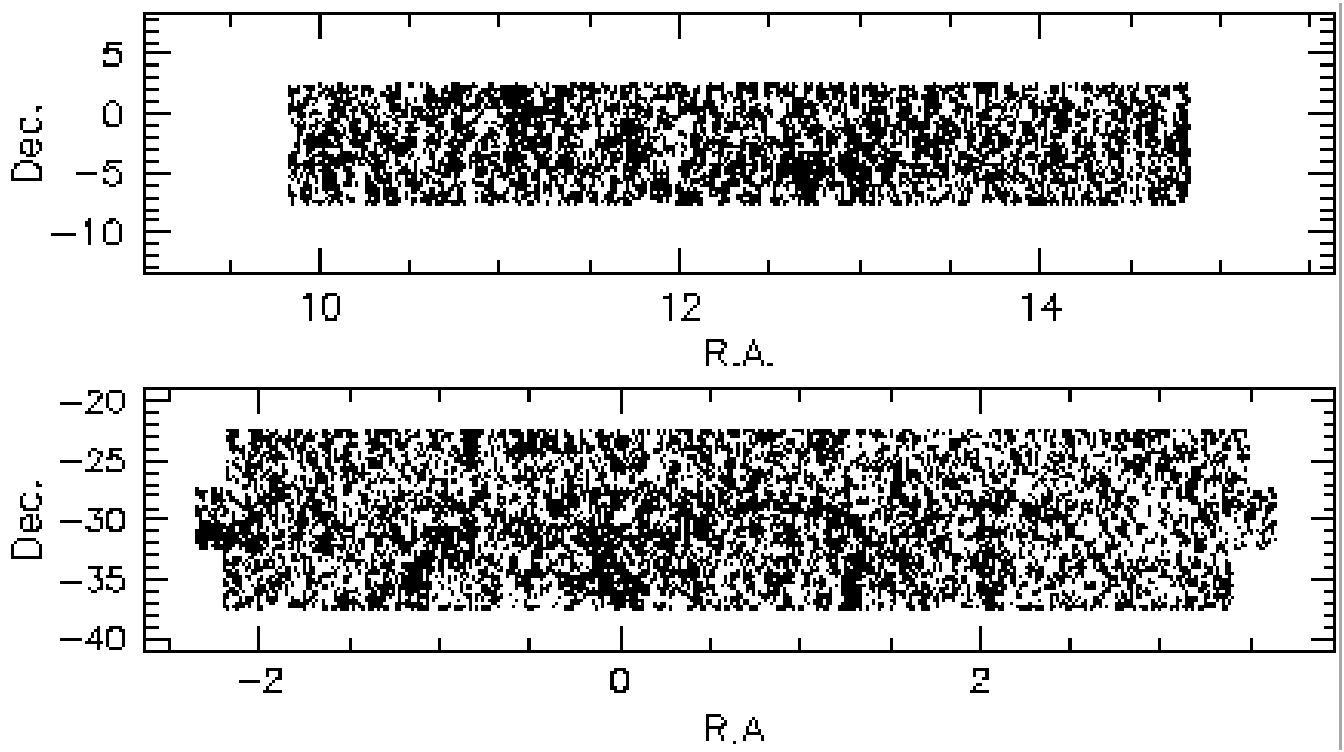}{1.0}
\caption{The distribution of the 7\% of objects in the sample that are
not allocated to fibres (NGP strip at top, SGP strip at bottom).}
\label{fig:unallocsky}
\end{figure*}

While the above procedure provides an optimised method of allocating as
many fibres as possible to a given individual field, it does not make
optimal use of the flexibility provided by the overlapping fields. We
therefore ran a second pass of the configuration algorithm which
considered each pair of overlapping fields as an 800 fibre problem in
which all objects found within the overlap area are matched up, so that
an object which is allocated in one field can be allocated to the second
field to allow acquisition of unallocated targets in the first field. 
Compared to the results of treating the two overlapping
fields as separate configuration problems, this procedure proves more
effective at reducing the overall incompleteness due to the
instrumental constraints. Since the tiling algorithm tends to place
local over-densities close to the edges of tiles, this procedure also
serves to homogenise the distribution of incompleteness as a function
of position within a field.

A significant fraction of this incompleteness arises due to close pairs
of objects which lie within the unique regions of individual fields. The
separation at which this occurs is a strong function of the relative
orientation of the two fibre buttons (see Lewis \etal\ 2001). The upper
limit on the value of this separation is 11mm in the focal plane
(corresponding to 2.2\arcmin), while the mean separation of close pairs
of objects for which one target cannot be assigned is $\sim$30\arcsec.

\subsection{Fibre assignment completeness}
\label{ssec:fibrecomplete}

The results of this procedure are that we are able to allocate fibres to
93\% of the source catalogue objects. The distribution of the unallocated
objects on the sky are shown in Figure~\ref{fig:unallocsky}. The most
prominent features visible in these distributions are occasional
localised clusters of unallocated objects. These are due to over-dense
regions where the geometrical packing constraints imposed by the fibre
button dimensions mean that it has not been possible to assign every
fibre to a target, even though there are enough fibres available as a
result of the tiling algorithm. This effect is enhanced by the relative
increase in the number of close pairs in strongly clustered regions.

\begin{figure}
\plotone{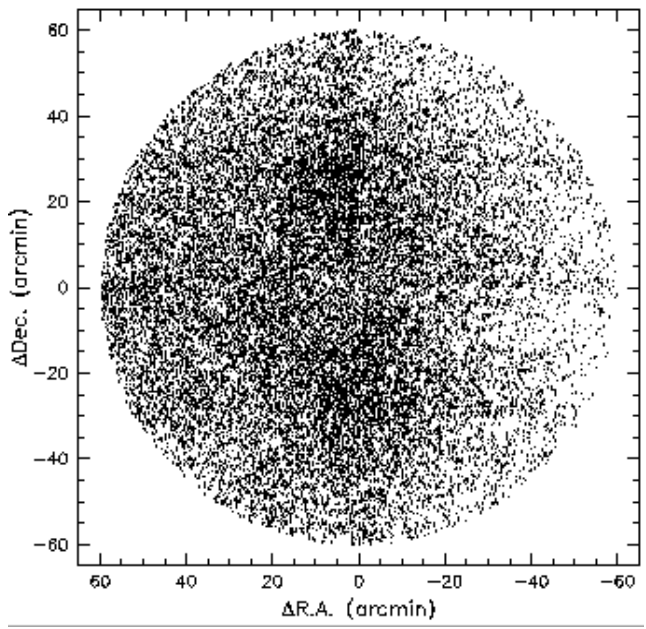}
\caption{The distribution of the unallocated objects within the stacked 2dF
fields.}
\label{fig:unallocplate}
\end{figure}

The distribution of these unallocated objects is also shown as a
function of position within the 2dF field in
Figure~\ref{fig:unallocplate}. This plot shows a slight overall gradient
in the R.A.\ direction, which is due to the order in which the field
overlaps are processed (see \S\ref{ssec:fieldoverlap}). This apparent
gradient is misleading, since the missed objects are plotted according
to the field plate coordinates on the plate to which they were assigned,
whereas the majority of objects which contribute to this effect are
actually missed on more than one 2dF field. The next most prominent
feature of Figure~\ref{fig:unallocplate} is the imprint of the four
fibres which are used for field acquisition and guiding. These fibres
must be allocated with a higher priority than the targets if the field
is to be observed successfully. They prevent the subsequent allocation
of target objects lying along their length, and so produce a slight
increase in the numbers of unallocated objects along the four cardinal
axes of the field, in the regions of the plate where they can be placed.

It should be noted that these figures do not represent the final
incompleteness in the survey, since individual target allocations are
adjusted immediately prior to observation to account for the actual
number of fibres available to 2dF on any given night (typically, about
1\% of the fibres are broken at any given time, though they are
continually replaced). The overall incompleteness at this stage, before
any observations have been made, is 7\%. The level of this
incompleteness is such that the pattern is dominated by residuals due to
clustering within individual two-degree fields, implying that an attempt
to correct for this effect would effectively imprint the observed
pattern onto the data. We therefore conclude that we have achieved our
goal of reducing the mean incompleteness to an acceptably low level,
comparable to the level of incompleteness due to the individual
uncertainties in galaxy magnitudes and the incompleteness due to not
measuring redshifts because of inadequate signal-to-noise. We emphasise
that neither the tiling nor the fibre allocation depend on the
photometric or spectroscopic properties of the objects, so that we can
use the source catalogue to accurately determine the sampling of the
survey, as discussed in \S\ref{sec:mask}).

\subsection{Visual inspection}
\label{ssec:inspection}

Postage stamp images of each target galaxy and each candidate fiducial
star were generated from the Digitized Sky Survey\footnote{The Digitized
Sky Survey was produced at the Space Telescope Science Institute, and is
based on photographic data obtained using the Oschin Schmidt Telescope
on Palomar Mountain and the UK Schmidt Telescope.} (DSS; Lasker \etal\
1998). These DSS images were used to examine the target galaxies for
merged images and to remove unsuitable fiducial star candidates.

The APM image classification parameter $k$ (see Maddox \etal\ 1990b) was
used to find target galaxies which might possibly be merged images.
Objects with $k$$>$1.15 in the SGP and $k$$>$1.2 in the NGP were
analysed by a simple routine which attempts to relocate the fibre onto
the local maximum of surface brightness closest to the nominal target
position. This is achieved by starting from the central pixel of the DSS
image and allowing the position to move in a direction of increasing
intensity until a local maximum is found. The position corresponding to
this local maximum is then inserted into the 2dF configuration file
prior to observation. All objects which are tested in this way are
flagged as merged images by changing the second character of the
object's {\tt OBSNAME} parameter from {\tt G} to {\tt M}, even if the
final position of the fibre is unchanged. The survey database records
both the original source catalogue position (database parameters {\tt
RA} and {\tt DEC}; see \S\ref{ssec:database})and the actual observed
position (database parameters {\tt OBSRA} and {\tt OBSDEC}); the
parameter {\tt MATCH\_DR} gives the offset (in arcsec) between these two
positions, and may be non-zero for objects flagged as mergers.

The DSS images of candidate fiducial stars are inspected visually to
allow removal of the following types of object, which would compromise
field acquisition: (i)~stars which are merged with another object;
(ii)~stars which are close to other stars of comparable brightness;
(iii)~bright stars which are strongly saturated on the sky survey
plates; (iv)~spurious objects that are fragments of the diffraction
spikes or halos of bright stars, or bright regions within large
galaxies; (v)~asteroids and portions of satellite trails; (vi)~spurious
objects due to noise.

\subsection{Final adjustments}
\label{ssec:adjustments}

The configuration procedure described above determines the optimal
assignment of fibres to targets over the whole survey. This procedure
assumes a fixed set of parameters for the transformation of
sky-coordinates to plate-coordinates, whereas in practice a slightly
different transform is determined each time the instrument is mounted on
the telescope. The above procedure also assumes a full complement of
fibres, and does not allow for the small but significant attrition rate
as fibres are disabled due to optical or mechanical failures. Fibres
which are disabled in this way are repaired on a regular basis, but
operational constraints require that disabled fibres are replaced in
batches, and so there is generally some delay before a particular fibre
is replaced.

We account for both the fibre attrition and the effects of changes to
the coordinate transform by re-examining each configuration the day
before that field is to be observed. All broken or inoperative fibres
which had been allocated are deallocated and the targets to which they
were assigned are flagged. An attempt is then made to recover each
flagged target using the current list of available fibres, according to
the same algorithm used for the original allocation. Again, low-priority
targets may be lost at this stage in favour of high-priority targets. In
practice, we usually find that the majority of the high-priority targets
which were flagged can either be recovered or replaced.

A similar check is then made to allow for a further change to the
coordinate transform due to the hour angle at which the field is to be
observed. This correction is a combination of the effects of atmospheric
refraction and flexure and misalignment of the telescope itself. Each
configuration is checked for hour angles of $\pm$4~hours, and
adjustments are made to ensure that no fibre conflicts will occur within
this range. Again, we usually find that all such conflicts can be
recovered without loss of target allocations.

After this correction to the configuration, a number of unallocated
fibres (at least 10 for each spectrograph) are allocated to blank sky
positions. This is done initially using an azimuthally symmetrical grid
of positions around the field, but if insufficient fibres can be
allocated to this grid, more are allocated to other blank sky regions
manually. The position of each allocated sky position is then checked to
ensure that it is genuinely blank by examining the surrounding area
using the Digitized Sky Survey. Any sky fibres likely to be contaminated
by objects visible on the DSS are reallocated to other blank sky
regions.

\section{SPECTROSCOPIC DATA}
\label{sec:data}

\subsection{Observational procedure}
\label{ssec:observing}

A target field is acquired using the four guide fibres, each of which
consists of 6 fibres in a hexagonal pattern surrounding a central fibre.
The individual fibres are 95$\mu$m (1.4\arcsec) in diameter and have
centre-to-centre separations of 1.8\arcsec. These guide fibres are
positioned on 14$<$$V$$<$15 stars with positions measured from the same
UKST plates as the target galaxies. As the blue sky survey plates are
often 20--25 years old, we use the more recent red sky survey plates to
check for any proper motion, and we also apply a colour cut to reduce
the number of nearby stars. On some fields, where red plates are not
available, we have used fiducial stars from the USNO A1.0 astrometric
catalogue (Monet \etal\ 1996).

The guide fibres are viewed by the AAT's standard acquisition and
guiding TV system. The position for optimum acquisition of the fiducial
stars, and subsequent guiding corrections, are determined by visual
inspection of the locations of the stars in the guide fibres on the TV
image. Acquisition of a new field typically requires less than
5~minutes.

For each field we first take a multi-fibre flat-field exposure using the
quartz lamp in the calibration unit. This flat-field is used to trace
the positions of the fibres on the image, to fit the spatial profile of
each fibre as a function of wavelength, and to apply a 1-dimensional
pixel-to-pixel flat-field correction to the extracted spectra. We next
take a wavelength calibration exposure of helium and copper-argon arc
lamps. We then typically take three 1100s exposures of the target
objects. However the exposure time is varied to suit the conditions: for
galaxy-only fields in good conditions, three 900s exposures are
sufficient; in poor conditions, especially for fields including targets
from the QSO survey, a longer total integration time is used. With the
CCD readout time of 60s, the whole series of exposures typically takes
just over an hour, so that the observations of one field are finished
just as the configuration for the next field is completed.

\subsection{Data reduction}
\label{ssec:reduction}

The data are reduced using the 2dF data reduction pipeline software,
\tdfdr, a full description of which is given by Bailey \etal\ (2001; see
also http://www.aao.gov.au/2df). The main steps in the process are as
follows:

 1. {\em Basic image processing.} Each image undergoes the same initial
    processing, which consists of flagging bad pixels (both saturated
    pixels and those specified in the bad-pixel mask for each CCD). The
    bias level is computed as the median of the CCD overscan region and
    subtracted from the image. The image is then trimmed to remove the
    overscan region. Finally, a variance array is computed based on the
    pixel data values and the known properties of the CCDs. The
    variances are propagated through all subsequent reduction steps.

 2. {\em Mapping the spectra.} The next step is to map the locations and
    shapes of the spectra on the detector. The centroids of the fibre
    spectra along the slit are determined using the fibre flat field
    image, and the fibre identifications checked visually. This is
    necessary because the positioning of the slit block is not precisely
    reproducible, so that the locations of the fibres on the CCD can
    change by a few pixels, and occasionally the first or last fibre may
    fall off the edge of the detector. Visual checking ensures that
    spectra are correctly identified with fibres and thus with the
    objects that were observed. Given this starting location, \tdfdr\
    finds the centroids of each fibre spectrum as a function of
    wavelength and fits a model based on the known optical properties of
    the spectrograph. For each subsequent frame, this model is adjusted
    (to account for flexure) by fitting an overall shift and rotation to
    the positions of the spectra on the detector.

 3. {\em Subtracting the background.} The background of scattered light
    is fitted with an empirically-determined model using the
    unilluminated portions of the detector. Fibres are so closely packed
    on the detector that the light level between transmitting fibres
    does not drop to the background level. Hence the only unilluminated
    areas are at the edges of the detector and at the positions of dead
    (broken, non-transmitting) fibres. A small number of fibres are
    broken at any one time and are continually being replaced.
    Occasionally the number of broken fibres becomes sufficiently small
    that it is not possible to reliably fit the background, and this
    step must be omitted.

 4. {\em Fitting the spatial profiles.} The close packing of the fibres
    on the detector means that their spatial profiles overlap, producing
    cross-talk between neighbouring spectra. In order to correct for
    this, it is necessary to know the spatial profile of each fibre
    spectrum as a function of wavelength. These profiles are determined
    from the fibre flat field image by performing simultaneous fits of
    200 Gaussians to the fibre profiles at each of 20 wavelength ranges.
    These fits are then interpolated in the wavelength direction with
    cubic splines to give the spatial profile for each fibre at every
    wavelength on the detector.

 5. {\em Spectrum extraction.} Given the location and shape of each
    fibre's spatial profile, optimal extraction consists of a weighted
    fit of the profiles to recover the amplitude of the spectrum at each
    wavelength. Because the spectra overlap, a simultaneous fit of the
    profile amplitudes of each fibre spectrum and its nearest neighbour
    on either side is carried out. The fit is performed using
    least-squares with variance weighting. 

 6. {\em Flat-fielding.} The extracted spectra are then `flat-fielded'
    by dividing by the pixel-to-pixel variations in the extracted
    spectra from the fibre flat field image. This one-dimensional
    flat-fielding is used because there is presently no means of
    generating a full two-dimensional flat-field image.

 7. {\em Wavelength calibration and linearisation.} A predicted
    wavelength for each pixel, based on the nominal central wavelength
    and the optical model for the spectrograph, is first calculated.
    Lines in the helium/argon arc spectra are automatically identified
    by a peak-finding algorithm, then matched to a list of known lines,
    discarding both unidentified peaks and unmatched lines, as as well
    as known blends. The relation between the measured line positions
    and their true wavelengths is fit by a third-order polynomial. This
    fit iterated up to four times, with the most discrepant line
    excluded at each iteration. The final fits usually include 21--22
    lines over the 4400\AA\ spectral range, and have typical rms
    residuals of 0.3\AA\ (0.07 pixels). Once the wavelength calibration
    is determined, the spectrum is re-binned onto a linear wavelength
    scale using quadratic interpolation.

 8. {\em Fibre throughput calibration.} The spectrum of the sky
    background is measured with at least 8 fibres (and usually more) on
    each of the two spectrographs. The spectra from the sky fibres are
    normalised by their mean fluxes and a median sky spectrum is
    calculated. The relative throughputs of the fibres are then derived
    from the relative fluxes in the strong sky lines as follows. All
    object and sky fibres and the median sky are continuum-subtracted
    using a continuum derived by median smoothing with a 201-pixel box.
    This removes the continuum leaving only the strong lines in the
    residual spectrum. A robust least-squares fit to the constant of
    proportionality between the counts in each pixel of the residual
    object+sky spectrum and the residual median sky spectrum then yields
    the relative throughput for each fibre (the sky lines in common to
    both spectra contribute to the fit, but strong emission lines found
    only in the object spectrum do not). The individual object and sky
    spectra are then normalised by dividing by their relative
    throughputs.

 9. {\em Sky subtraction.} The median sky spectrum is re-calculated from
    the individual normalised sky spectra, and subtracted from each
    normalised object and sky spectrum. The sky-subtraction precision,
    measured as the ratio of the total flux in the sky fibres before and
    after sky-subtraction, is typically 2--3\% and rarely worse than
    6\%. Precise sky-subtraction is dependent on effective scattered
    light subtraction, wavelength calibration and throughput
    calibration.

10. {\em Combining spectra.} The fully-reduced spectra from each of the
    multiple exposures on the target objects (typically three) are
    optimally combined in a final step. The combination algorithm
    accounts for overall flux variations between exposures (due to
    different integration times or observing conditions) and rejects
    cosmic ray events. The relative weighting of the exposures is
    computed from the total fluxes in the brightest 5\% of the spectra
    (after they are median-smoothed over a 201-pixel box to eliminate
    cosmic rays and normalised to the first exposure). The maximum flux
    weight is unity, and the weights are applied to the individual
    exposures before they compared for cosmic rays. The cosmic ray
    rejection algorithm is similar to that of the IRAF {\tt crreject}
    package. It flags likely cosmic ray events, which are taken to be
    those pixels that are more than 5$\sigma$ deviations from the median
    (over the set of flux-weighted exposures) together with some
    neighbouring pixels. The final pixel value in the combined spectrum
    is computed as the variance-weighted mean over the flux-weighted
    exposures of the unflagged pixels.

The final result of the reduction is a pair of data files (one for each
CCD), both of which contain a multi-spectrum image (200 fibres $\times$
1024 spectral pixels), the associated variance array (which is correctly
propagated throughout the reduction process), and a copy of the median
sky spectrum used in the background subtraction. The data files also
contain the identifications of the spectra obtained from the fibre
configuration file and detailed information about the instrument and
observation.

The spectra are not flux-calibrated. Flux calibration of fibre spectra
is very difficult to achieve for a number of reasons. Chief among these
are: (i)~astrometric/positioning errors, which, combined with the small
fibre aperture, prevent reproducible sampling of each galaxy's light
distribution; (ii)~chromatic variations in distortion (see
\S\ref{sec:instrument}) and residual errors in the atmospheric
dispersion correction, which produce variations in the flux calibration
dependent on the attitude of the telescope and the position of the
object in the field. These effects can produce significant variations in
the continuum slope of the spectra, and are discussed in quantitative
detail in Madgwick \etal\ (2001b) and Madgwick \etal\ (2001c, in
preparation).

Various problems are exhibited by some of the spectra: (i)~{\em Poor
sky-subtraction.} A small but significant fraction of fields suffered
sky-subtraction errors of more than 5\%. The strong oxygen sky lines at
5577\AA, 6300\AA, 6363\AA\ and the OH bands to the red of 6000\AA\ leave
significant residuals in the spectra. (ii)~{\em Fringing.} Damaged
fibres sometimes produce a fringing effect that results in a strong
oscillation in the throughput with wavelength. (iii)~{\em Halation.}
Sometimes halation due to condensation on the field flattener lens just
in front of the CCD results in a high level of scattered light which
fills in absorption features; when present, this effect was worse in CCD
camera \#2. (iv)~{\em ADC error.} A software error meant that the
atmospheric dispersion corrector (ADC) was incorrectly positioned for
observations prior to 31 August 1999; some of the spectra obtained
before this date suffer from significant atmospheric dispersion losses.

\section{REDSHIFTS}
\label{sec:redshifts}

\subsection{Redshift estimation}
\label{ssec:zcode}

The main parameter to be determined from the spectra was the redshift of
each object. A sophisticated and highly-tuned redshift code was
developed in order to achieve a high level of precision, reliability and
automation in the measurement of redshifts from the 2dFGRS spectral
data.

The redshift code (version 990505) uses two quasi-independent automated
methods: `absorption' redshifts are obtained by cross-correlation with
template spectra (after clipping emission lines) and `emission'
redshifts are obtained by finding and fitting emission lines. These
automated redshift estimates are followed by a visual check and
(occasionally, where necessary) a `manual' redshift obtained by fitting
identified spectral features.

Before these redshift estimates are made, however, the spectra are
pre-processed in a number of ways. Residual features from the strong
atmospheric emission lines at 5577\AA, 5893\AA, 6300\AA\ and 7244\AA\
are masked by interpolating over a small wavelength range (20--30\AA)
centred on the sky line. The atmospheric absorption bands around
6870\AA\ (B-band) and 7600\AA\ (A-band), and the fibre absorption band
around 7190\AA, are removed by dividing by an approximate band
correction. This correction is obtained by fitting a low-order
polynomial to the mean of all the spectra in the reduced image, dividing
the mean spectrum by this smooth fit, finding the region about each band
centre where this ratio is less than unity, and setting the band
correction to the value of this ratio in these regions and to unity
elsewhere. The spectra, which are not flux-calibrated, are finally
multiplied by a simple quadratic approximation to the mean flux
calibration correction in order to give appropriate weighting across the
whole spectral range.

\begin{figure}
\plotone{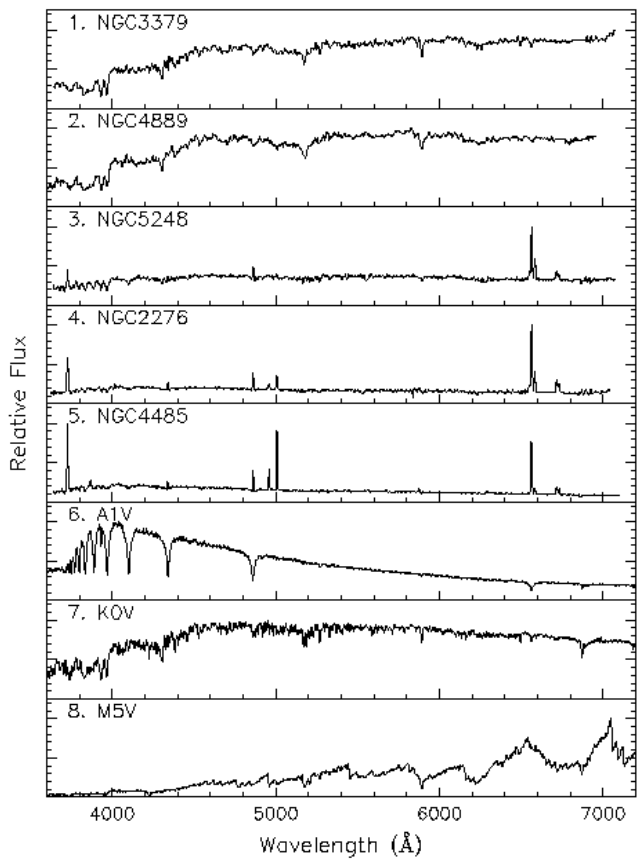}
\caption{The eight template spectra against which the 2dFGRS galaxy
spectra were cross-correlated.}
\label{fig:tempspec}
\end{figure}

An `absorption' redshift is obtained by cross-correlating the galaxy
spectrum against a suite of 8 template spectra following the general
method of Tonry \& Davis (1979). The templates that are used cover a
broad range of spectral types, and include 5 galaxies and 3 stars. The
galaxies (and their morphological types) are: 1.~NGC3379 (E),
2.~NGC\,4889 (cD), 3.~NGC\,5248 (Sbc), 4.~NGC\,2276 (Sc) and
5.~NGC\,4485 (Sm/Im); the stars (and their spectral types) are:
6.~HD\,116608 (A1V), 7.~HD\,23524 (K0V) and 8.~BD\,05\,1668 (M5V). The
galaxy spectra come from spectrophotometric atlas of Kennicutt (1992);
the stellar spectra come from the library of Jacoby \etal\ (1984). The
full set of template spectra are shown in Figure~\ref{fig:tempspec}.

The galaxy and template spectra are prepared for cross-correlation using
the following steps: (1)~continuum-subtraction using a sixth-order
polynomial fit to the continuum; (2)~removal of strong emission lines by
patching out all points more than 5 times the rms variation above the
mean; (3)~re-binning to a logarithmic wavelength scale with 2048 pixels;
(4) apodising (tapering the ends of the spectrum to zero) by multiplying
the first and last 5\% of the spectrum with a cosine bell; (5)~Fourier
transformation; and, finally, (6)~multiplication by an exponential
filter of the form $e^{-k/k_h}-e^{-k/k_l}$, where $k$ is wavenumber in
inverse pixels, $k_h$=300\perpix\ and $k_l$=15\perpix, in order to
reduce the effects of both the residual continuum at low wavenumber and
the noise at high wavenumber.

The cross-correlation function is then computed as the inverse Fourier
transform of the complex product of the filter Fourier transforms of the
processed galaxy and template spectra. The highest peak in
cross-correlation function is fitted with a quadratic in order to obtain
the position and height of the peak. The significance of the peak is
measured in terms of the ratio, $R$, of the height of the peak to the
noise in the cross-correlation function, which is estimated from its
anti-symmetric part. The absorption redshift is taken to be the
cross-correlation redshift obtained with the template giving the highest
value of $R$. An estimate of the quality of the absorption redshift,
Q$_a$, is based on the value of $R$, with Q$_a$=1,2,3,4 for
$R$$>$3.5,4.0,4.5,5.0 and Q$_a$=0 otherwise (with the additional
requirement that Q$_a$=3 and Q$_a$=4 also require, respectively, at
least 4 and 6 of the 8 templates give the same redshift to within
600\kms).

An `emission' redshift is obtained by fitting Gaussians to significant
spectral features and searching for a multi-line match. The galaxy
spectrum is first continuum-subtracted using a sixth-order polynomial
fit to the continuum and lightly smoothed with a Gaussian kernel having
a dispersion of 0.8~pixels. Peaks in the smoothed spectrum above a
threshold (set at 3.3 times the robustly-estimated rms variation) are
flagged as candidate emission lines. For each of these, in descending
order of strength, a Gaussian is then fitted to the unsmoothed data,
using the peak in the smoothed spectrum as the initial guess for the
line centre. The Gaussian fits to the stronger lines are subtracted in
turn before the weaker lines are fitted (to minimize blending effects in
the H$\alpha$/NII pair). Then lines with fitted FWHM between 0.7 and 7
pixels, and total `significance' of the whole line above 3.5$\sigma$ are
marked as `good'; this rejects unclipped cosmic rays and
low-significance features. 

The `good' lines are then sorted by strength, and the three strongest
tested to see if at least two can be identified as common emission lines
([OII]\,3727\AA, H$\beta$\,4861\AA, [OIII]\,5007\AA, H$\alpha$\,6563\AA\
or [NII]\,6584\AA) at a single redshift (the two out of three criterion
is used to allow for one spurious feature, such as a residual sky line
or cosmic ray). If so, then all the other lines in this set (plus
[OIII]\,4959\AA) that match this redshift to better than 600\kms\ are
found and the mean redshift from these matching lines is adopted as the
emission redshift. If not, then up to two possible single-line redshifts
are kept for comparison with the absorption redshift, assuming the
single line is either [OII] or H$\alpha$ in the range 0$<$$z$$<$0.4, as
appropriate. An estimate of the quality of the emission redshift, Q$_e$,
is based on the number and strength of the lines from which it is
determined, with Q$_e$=0 if there are no lines, Q$_e$=1 for one weak
line, Q$_e$=2 for any two lines or one strong line, and Q$_e$=4 for
three or more lines.

The penultimate step is the choice of the best automated redshift
estimate, given the absorption and emission redshifts and their quality
parameters. The best redshift is taken to be whichever of the absorption
and emission redshifts has the higher quality parameter (with the
absorption redshift preferred if Q$_a$=Q$_e$). The best estimate of the
redshift quality is Q$_b$\,=\,max(Q$_a$,Q$_e$), with two exceptions:
(i)~if the difference between the absorption and emission redshifts is
less than 600\kms\ then Q$_b$\,=\,max(Q$_a$,Q$_e$,3); and (ii)~if
Q$_a$$\ge$2 and Q$_e$$\ge$2 and the difference between the absorption
and emission redshifts is greater than 600\kms, then this discrepancy is
flagged and we set Q$_b$=1.

The final step in measuring the redshift is a visual check of the best
automatic redshift. The redshift code displays for the user both the
cross-correlation function and the galaxy spectrum with all the common
spectral features superposed at the best automatic redshift. The
spectrum displayed is the version actually used for the redshift
estimate, with bad pixels and the strongest night sky lines patched out,
and with an approximate correction applied for the instrumental
response. To assist in identifying spurious features the code also
displays the mean sky spectrum (to show the positions of the sky
features), the atmospheric absorption bands and the variance array
associated with the spectrum. The user is given the emission and/or
absorption line redshifts, the best-estimate redshift, whether this is
an emission or absorption redshift (or both), and whether (if both are
obtained) the emission and absorption redshifts agree or not. If neither
of the automated estimates are correct, the user can manually estimate a
redshift by identifying a particular spectral feature and fitting
Gaussians at the positions of all the common features. These fits can be
to emission or absorption features and are rejected if there is no
detectable feature near the nominal position. The manual redshift is
taken to be the mean of the redshifts estimated from each of the
well-fitted features. Given all this information, the observer can
visually inspect the spectrum, try alternative redshifts (absorption,
emission and manual) and determine the best redshift estimate.

Once the final redshift estimate has been decided, the user assigns a
quality, Q, to this redshift on a five-point scale, with the nominal
interpretation (see below for further discussion): Q=1 means no redshift
could be estimated; Q=2 means a possible, but doubtful, redshift
estimate; Q=3 means a `probable' redshift (notionally 90\% confidence);
Q=4 means a reliable redshift (notionally 99\% confidence); Q=5 means a
reliable redshift and a high-quality spectrum. Note that this quality
parameter is determined entirely by the subjective judgement of the
user, and is independent of the automatic quality parameter Q$_b$.
Quality classes 1 and 2 are considered failures, so the redshift
completeness is the number of Q=3,4,5 redshifts divided by the total
number of galaxies in the field. The standard redshift sample comprises
objects with Q$\ge$3, but (for applications where redshift reliability
is more important than completeness) one may prefer to use the set of
objects with Q$\ge$4.

For the objects ending up with `acceptable' (Q=3,4,5) redshifts, the
best automated redshift is the same as the best visual redshift about
93\% of the time; about 5\% of the time either the absorption or the
emission redshift agrees with the best visual redshift but the code has
preferred the other one; and about 2\% of the time neither of the
estimates matches the best visual redshift, and a manual redshift
estimate is required.

While there is some element of subjectivity in deciding whether a
redshift estimate is `acceptable' or not, this visual assessment is
found to improve the reliability of the final sample, since the observer
can assess a number of factors that are hard to fully automate, such as
the overall shape of the spectrum, line ratios, the general quality of
sky subtraction, and the proximity of features to sky lines or
absorption bands. For 15,000 spectra, two or more users have measured
redshifts independently using the same redshift code on the same
spectra; the pair-wise `blunder' rate (see below) is 0.4\%, which is
several times smaller than that for different observations of the same
object. The rms difference in redshift completeness (fraction of objects
per field with Q$\ge$3) between two users for the same data is less than
half of that between the {\em same} observer measuring redshifts for
many different fields. The human variation is therefore substantially
smaller than the unavoidable variation between fields due to weather and
other effects, and will to first order be corrected by the completeness
factors.

\subsection{Quality of spectra}
\label{ssec:quality}

\begin{figure}
\plotone{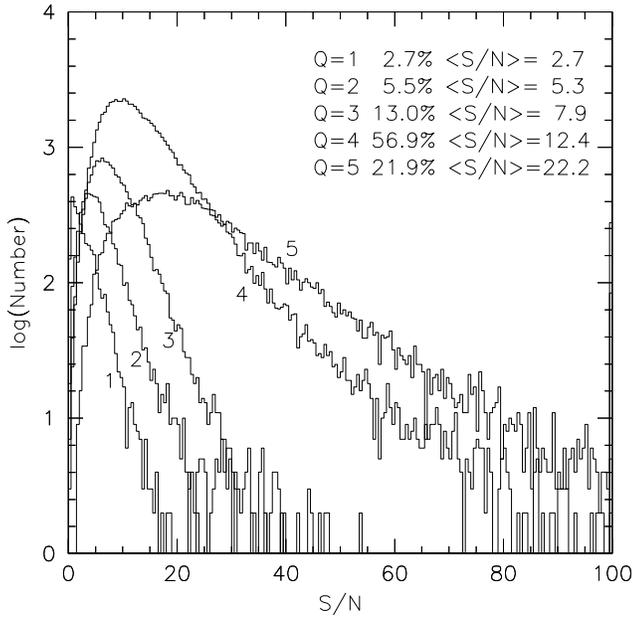}
\caption{The distribution of the median S/N per pixel for objects in
each of the 5 quality classes.}
\label{fig:snr}
\end{figure}

A quantitative overall measure of spectral quality is the median
signal-to-noise ratio (S/N) per pixel computed over all pixels in the
wavelength range 4000--7500\AA\ with fluxes greater than zero. This is
computed directly from the spectrum and it associated variance array,
and can be compared to the quality class Q assigned to the redshift
identification. The distribution of S/N over the 5 quality classes for
the best spectrum of each object (i.e.\ the spectrum on which the
adopted redshift is based) is shown in Figure~\ref{fig:snr}. The
fraction of objects in each quality class and the median S/N for each
quality class are given in the figure legend. Objects with reliable
redshifts (Q=3,4,5) make up 91.8\% of the sample and have a median S/N
of 13.1\perpix\ (6.3\,\AA$^{-1}$).

\begin{figure}
\plotone{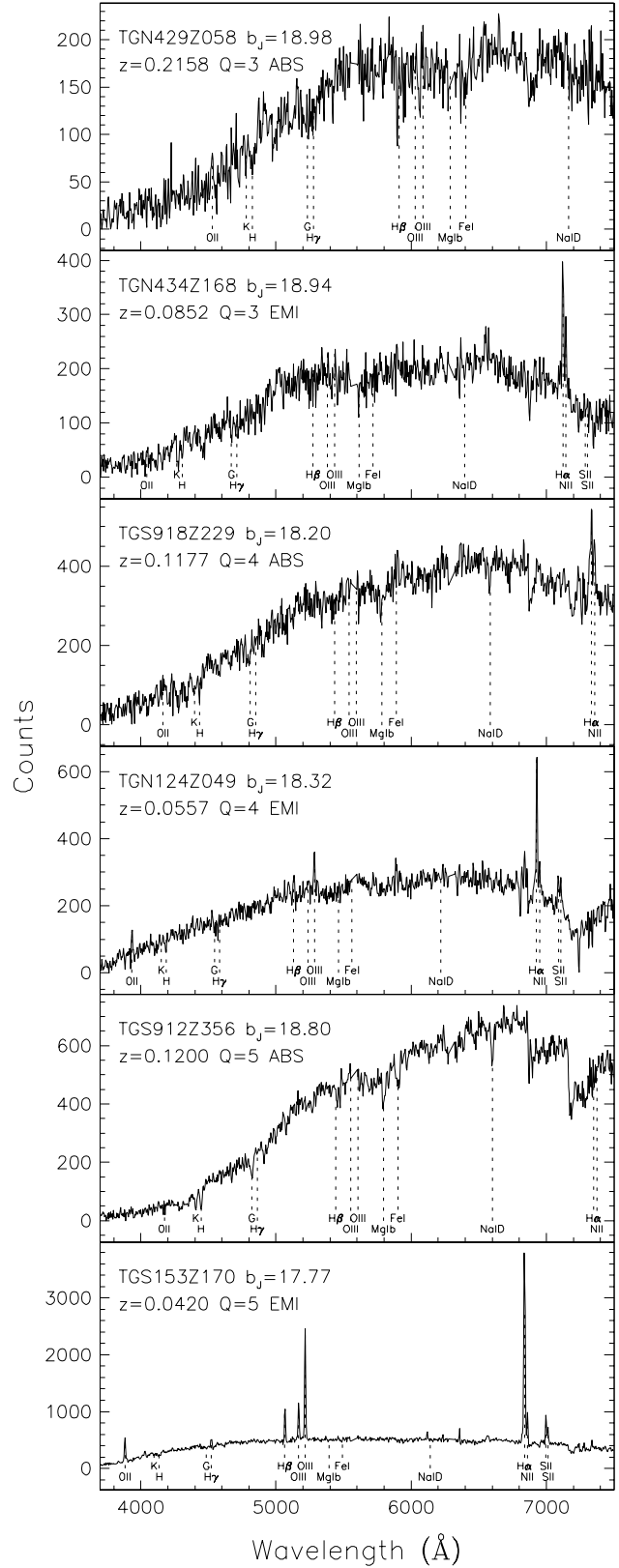}
\caption{Example 2dFGRS spectra, showing objects with quality classes
Q=3, Q=4 and Q=5, and both absorption (ABS) and emission (EMI)
redshifts. The most common spectral features are marked at the
wavelengths corresponding to the measured redshift of each object.}
\label{fig:galspec}
\end{figure}

Example 2dFGRS spectra for quality classes Q=3, Q=4 and Q=5 are shown in
Figure~\ref{fig:galspec}; for each quality class, one object with an
absorption redshift and one object with an emission redshift are shown.
In order to be representative, the spectra are chosen to have a S/N per
pixel close to the median of their quality class. Residuals from the
strong night sky lines at 5577\AA\ and 6300\AA\ have been interpolated
over.

\subsection{Repeat observations}
\label{ssec:intred}

Amongst the first 150\,000 galaxies in the survey with redshifts, there
are 8480 repeat redshift measurements; i.e.\ a second, or even a third,
reliable (Q=3,4,5) redshift measurement for an object. These repeat
measurements are either objects that lie in the overlaps between fields
(7444 cases) or are in fields which were observed more than once,
usually because the first observation was of poor quality and low
completeness (1036 cases). For these objects we can compare their best
redshift measurement with the repeat measurements to estimate both the
rate of incorrect redshift identifications in the survey and the
uncertainties on the correct redshift identifications.

\begin{figure}
\plotone{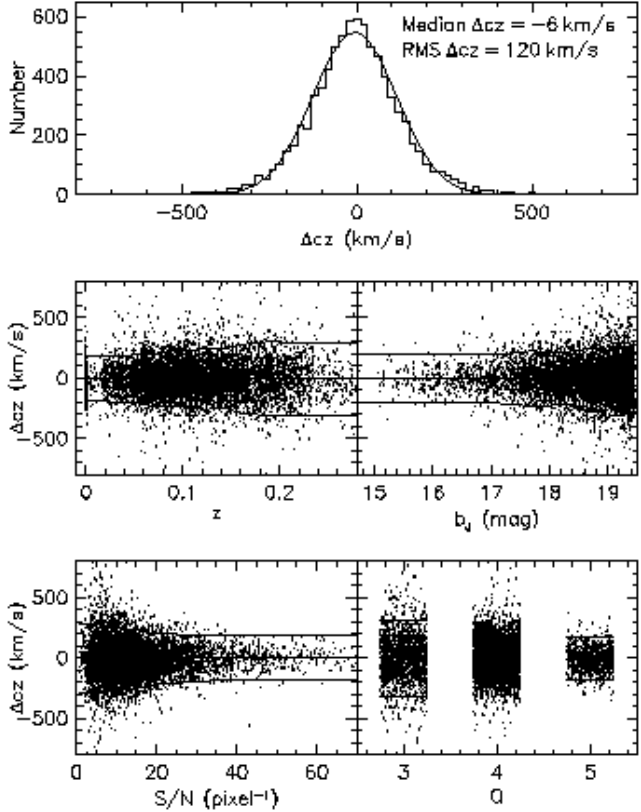}
\caption{The distribution of redshift differences between repeat
measurements with Q$\ge$3 is shown in the top panel, along with the
Gaussian having the same median and rms. The dependence of the redshift
differences on redshift, apparent magnitude, signal-to-noise ratio and
quality class for the best observation, is shown in the lower panels
(note that the quality classes have small random offsets added to make
the distribution of redshift differences clear). The lines in the lower
panels are the running median and the $\pm$2$\sigma$ ranges.}
\label{fig:repeats}
\end{figure}

The top panel of Figure~\ref{fig:repeats} shows the distribution of the
redshift differences (w.r.t.\ the best redshift measurement for the
object) for all the repeat observations. The rms redshift difference
(robustly estimated here and throughout as the half-width of the central
68\% of the distribution) is 120\kms, implying an overall
single-measurement rms uncertainty of 85\kms.

The lower panels of the figure show the trends in the redshift
uncertainty as a function of redshift, magnitude, S/N and quality class
Q. As expected, the uncertainties increase at higher redshifts (from
95\kms\ for $z$$<$0.05 to 150\kms\ at $z$$>$0.15), fainter magnitudes
(from 100\kms\ for \bj$<$18.5 to 140\kms\ for \bj$>$19.3) and lower S/N
(from 95\kms\ for S/N$>$25 to 145\kms\ for S/N$<$5). The strongest trends
are with redshift quality class and the method of estimating the
redshift. Table~~\ref{tab:repeats} gives the rms precision of single
measurements as function of redshift quality class or measurement method
(ABS = cross-correlation of absorption features; EMI = automatic fit to
emission lines; MAN = manual fit to features). The fact that manual
redshift estimation has the largest uncertainties mainly reflects the
fact that it is only employed when the automatic methods have failed,
usually in cases of low S/N.

\begin{table}
\centering
\caption{Summary statistics by quality class and method.}
\label{tab:repeats}
\begin{tabular}{ccccccccc}
Quality  & Fraction  & RMS error & Blunder \\ 
class    & of sample & (\kms)    & rate    \vspace{6pt} \\
Q=1    &   ~2.5\%  & ---       & ---     \\
Q=2    &   ~4.8\%  & 143       & 69.5\%  \\ 
Q=3    &   11.9\%  & 123       & 10.2\%  \\ 
Q=4    &   57.3\%  & ~89       & ~0.9\%  \\ 
Q=5    &   23.5\%  & ~64       & ~0.2\%  \\ 
         &           &           &         \\
Redshift & Fraction  & RMS error & Blunder \\
method   & of sample & (\kms)    & rate    \vspace{6pt} \\
Q=1,2  &   ~7.3\%  & ---       & ---     \\
ABS      &   72.0\%  & ~87       & ~0.7\%  \\
EMI      &   18.3\%  & ~61       & ~0.9\%  \\
MAN      &   ~2.4\%  & 159       & ~9.3\%  \\
\end{tabular}
\end{table}

We divide the repeat measurements into two categories: correct
identifications, where the redshift difference is less than 600\kms, and
`blunders', where it is greater than 600\kms. The division at 600\kms\
corresponds to 5 times the rms error in the redshift differences. It
should be noted that these blunders are not necessarily due to incorrect
redshift measurements for one (or both) of the observed spectra. They
may also be due to two observations of a close pair of objects with
slight offsets in the fibre position, resulting in a different member of
the pair dominating the spectral flux in each observation.

\begin{figure}
\plotone{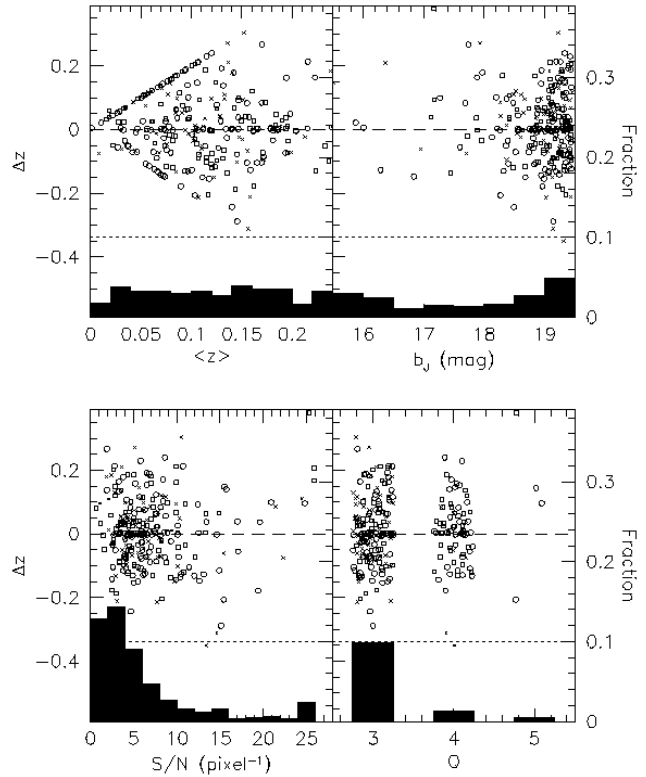}
\caption{The distribution of the redshift differences for the blunders
(repeat observations with $|\Delta cz|>600\kms$) against redshift,
magnitude, S/N and quality class. The circles are absorption redshifts,
the squares are emission redshifts and the crosses are manual redshifts.
The histograms at the foot of each panel show the fraction of blunders
as a function of redshift, magnitude, S/N and quality class.}
\label{fig:rejects}
\end{figure}

There are 263 blunders, so that the pair-wise blunder rate amongst the
Q=3,4,5 repeat measurements is 3.1\%. This implies a
single-measurement blunder rate of just 1.6\%. Figure~\ref{fig:rejects}
shows plots of the redshift differences for the blunders with respect to
redshift, magnitude, S/N and quality class; it also shows the pair-wise
fraction of blunders as a function of each of these quantities.

There is no increase in the blunder rate with redshift, but a
significant increase at fainter magnitudes, lower S/N and lower quality
class. The single-measurement blunder rates as functions of quality
class and measurement method are given in Table~\ref{tab:repeats}. We
find that Q=3 corresponds to individual redshifts having 90\%
reliability, and that Q=4 and~5 correspond to better than 99\%
reliability, as intended. The blunder rate for Q=2 redshifts is 70\%,
consistent with the definition of Q=2 as `possible but doubtful'. The
blunder rates for the absorption and emission redshifts are less than
1\%, but the blunder rate for manual redshifts is 9\% (due to the fact
that manual redshifts are only used in the 2.4\% of cases where the
automatic methods fail).

The distribution of blunders shows a slight peak around zero redshift
difference, suggesting that the tail of the error distribution for
correct redshift identifications is more extended than a Gaussian. Some
of the blunders may thus be correct identifications with larger than
normal errors, implying that we may be slightly over-estimating the
blunder rate. However, the most notable feature in the distribution of
blunders is the large number of instances (22\% of the total) where one
or other of the redshifts in the comparison is approximately zero. The
redshift difference is defined as (other redshift $-$ best redshift), so
the upper envelope corresponds to objects where the best redshift is
zero and the lower envelope corresponds to the other redshift being
zero. In the former case the best redshift indicates that the objects
are stars and the other (non-zero) redshift is due to a blunder based on
a poor-quality spectrum; in the latter case we have the reverse of this
situation. There are 3 times as many objects in the upper envelope as
the lower, so stars being mistaken for galaxies is a much more common
error than galaxies being mistaken for stars---which is remarkable given
that galaxies are 20 times more common than stars in our sample.

We have repeated the above analyses after separating the repeat
observations into those cases where the repeat is due to re-observing a
field (12\%) and those where it is due to observing an object in
overlapping fields (88\%). We find no significant difference in either
the single-measurement rms precision or the blunder rate for these two
subsets of repeat measurements.

Examination of the individual object spectra in the cases where there is
a blunder shows that in general the best redshift appears to be quite
reliable and that the main reason for the discrepancy is that the other
redshift measurement has been obtained from a poor-quality (low-Q)
spectrum.

\subsection{External comparisons}
\label{ssec:extred}

In order to compare our redshift measurements with those from other
sources, we have matched our catalogue to the redshift catalogues from
the second Center for Astrophysics redshift survey (CfA2; Huchra \etal\
1999), the Stromlo-APM redshift survey (SAPM; Loveday \etal\ 1996), the
PSCz survey (PSCz; Saunders \etal\ 2000), the Las Campanas redshift
survey (LCRS; Shectman \etal\ 1996), and also to the heterogeneous
redshift compilation of the ZCAT catalogue (version November 13, 2000;
J.P.\ Huchra, priv.comm.). Objects in these catalogues were matched to
objects in the 2dFGRS by position; positions separated by 4~arcsec or
less were assumed to belong to the same object. This criterion was
chosen to include most genuine matches and exclude most false matches
between close but unrelated objects.

\begin{figure}
\plotone{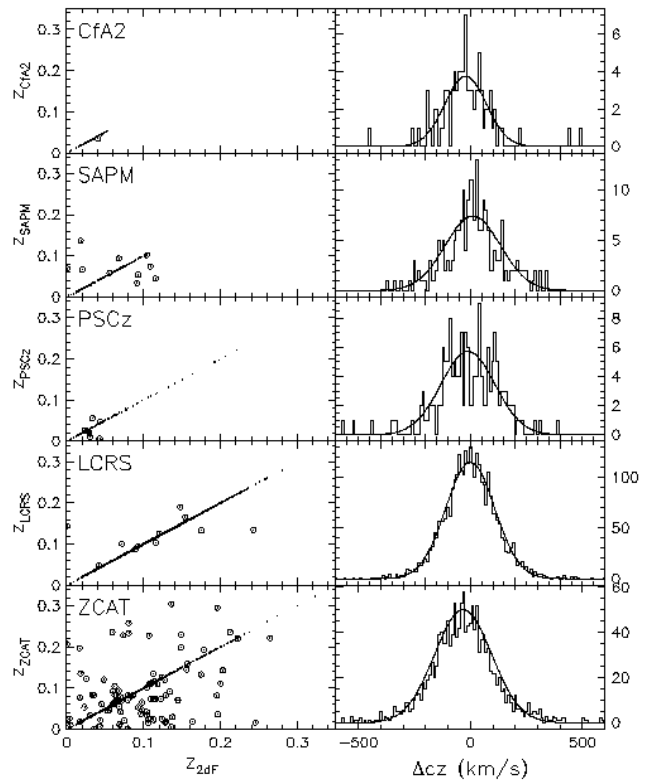}
\caption{Comparisons of the 2dFGRS redshifts with those from the CfA2,
SAPM, LCRS and ZCAT catalogues. The left panels shown the correlations
between the redshift measurements, with each comparison shown as a dot.
Most dots lie along the 1-to-1 line; the relatively few blunders
($|\Delta cz|>600\kms$) are circled. The right panels show the histogram
of redshift differences overlaid with a Gaussian having the same median
and rms.}
\label{fig:extred}
\end{figure}

Table~\ref{tab:extred} and Figure~\ref{fig:extred} summarise the results
of these comparisons. The table lists, for each external source
catalogue: the number of matched objects in the comparison, $N_{comp}$;
the number of cases, $N_{600}$, where the redshift difference between
the 2dFGRS and the other source indicates a blunder ($|\Delta
cz|>600\kms$); the median redshift difference, $\langle \Delta cz
\rangle$; and the rms scatter of $\Delta cz$ about this median,
$\sigma(\Delta cz)$. The median and rms scatter are estimated excluding
the blunders.

\begin{table}
\centering
\caption{Comparisons of redshifts with external sources.}
\label{tab:extred}
\begin{tabular}{lrrrr}
Source & $N_{comp}$ & $N_{600}$ & $\langle \Delta cz \rangle$ & $\sigma(\Delta cz)$ \\
       &            &           & \kms                        & \kms \vspace{6pt} \\
CfA2 &   87 &   1 & $-$22 &  93 \\
SAPM &  228 &  10 & $+$11 & 123 \\
PSCz &  172 &   7 & $-$11 & 120 \\
LCRS & 3135 &  11 &  $-$1 & 109 \\
ZCAT & 1593 & 103 & $-$33 & 127 \\
\end{tabular}
\end{table}

\begin{figure}
\plotone{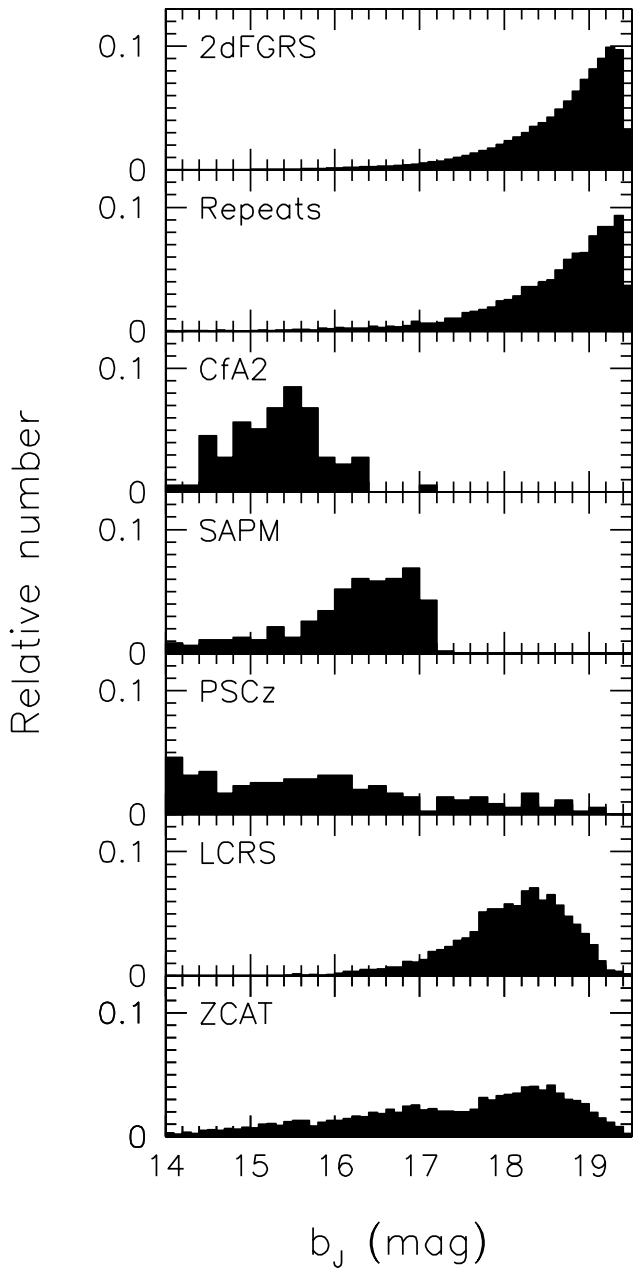}
\caption{The magnitude distributions of the 2dFGRS sample, the repeat
observations in the 2dFGRS sample, and the five samples of objects in
common with other redshift catalogues: CfA2, SAPM, PSCz, LCRS and ZCAT.}
\label{fig:magcomp}
\end{figure}

In interpreting these comparisons we need to keep in mind the magnitude
range of the objects that are in common. Figure~\ref{fig:magcomp} shows
the magnitude distributions of the comparison samples, all of which are
significantly brighter than the 2dFGRS itself. The CfA2, SAPM and PSCz
catalogues only overlap with the bright tail, while the median magnitude
of the LCRS is a magnitude brighter than that of the 2dFGRS. ZCAT has a
broader magnitude range, but is also heavily weighted towards brighter
objects.

The best comparison is with the LCRS: the comparison sample is the
largest, the magnitude range most similar, and the catalogue is
homogeneous (unlike ZCAT). The rms redshift difference of 109\kms\ is
consistent with the single-measurement uncertainty of 76\kms\ for
objects in the 2dFGRS brighter than \bj=18.5 and the typical uncertainty
of 67\kms\ on the LCRS redshifts (Shectman \etal\ 1996). The pair-wise
blunder rate in the comparison sample is 11/3135 = 0.4\%, which in fact
is rather lower than might be expected from the pair-wise blunder rate
of 46/2920 = 1.6\% for repeat measurements of objects in the 2dFGRS
brighter than \bj=18.5; this may be due to the high surface brightness
selection criterion of the LCRS.

The worst comparison is with ZCAT: the rms redshift difference is
127\kms\ and the pair-wise blunder rate is 103/1593 = 6.5\%, which is
more than twice the pair-wise blunder rate for the 2dFGRS repeats. Given
the single-measurement blunder rate for the 2dFGRS is 1.6\%, the implied
single-measurement blunder rate for ZCAT is 5.0\%, which is presumably
due to the heterogeneous nature of its source catalogues.

Although the samples in common with CfA2, SAPM and PSCz are relatively
small, they give results that are consistent with the rms redshift
errors and pair-wise blunder rate obtained from the 2dFGRS repeat
observations. 

Checking the 2dFGRS spectra of the objects that have redshift blunders,
we find that 0/1 cases in the comparison with CfA2, 1/10 for SAPM, 1/7
for PSCz and 4/11 for LCRS might {\em possibly} be attributed to errors
in the 2dFGRS redshift estimates. Examining the DSS images of these
objects, we find that 2 of the discrepancies with SAPM, 1 of those with
PSCz, and 6 of those with LCRS, occur in cases where there is a close
companion to the target object and hence some potential for confusion.

\section{SURVEY MASKS}
\label{sec:mask}

\begin{figure*}
\plotfull{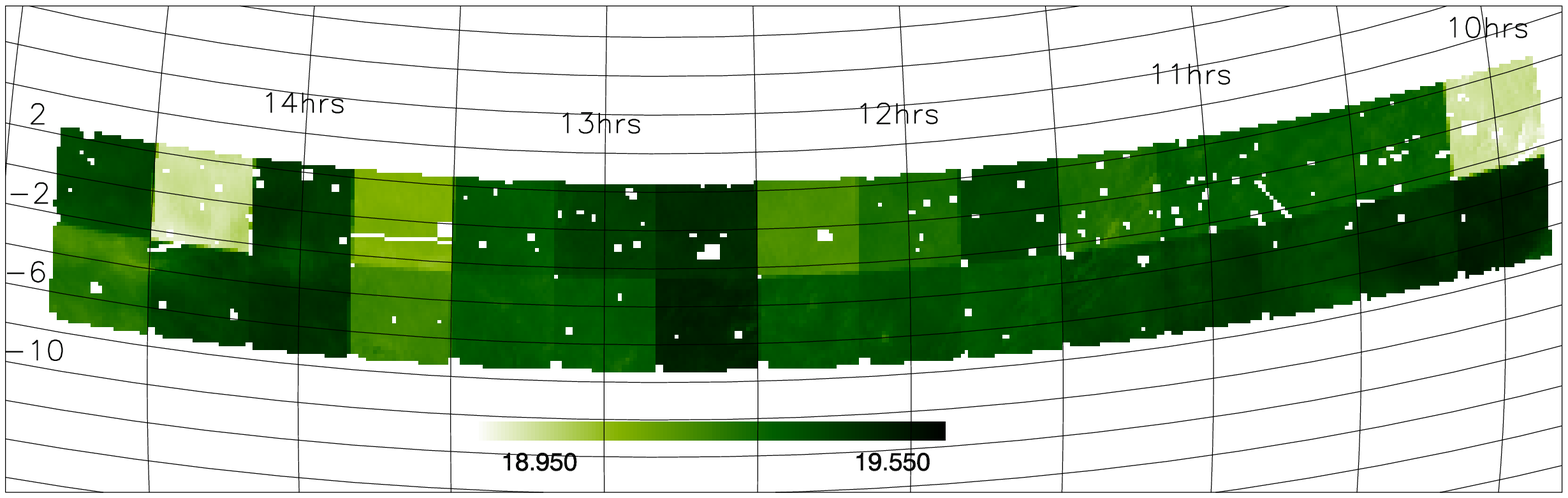}{1.0} \vspace{1pt} \\
\plotfull{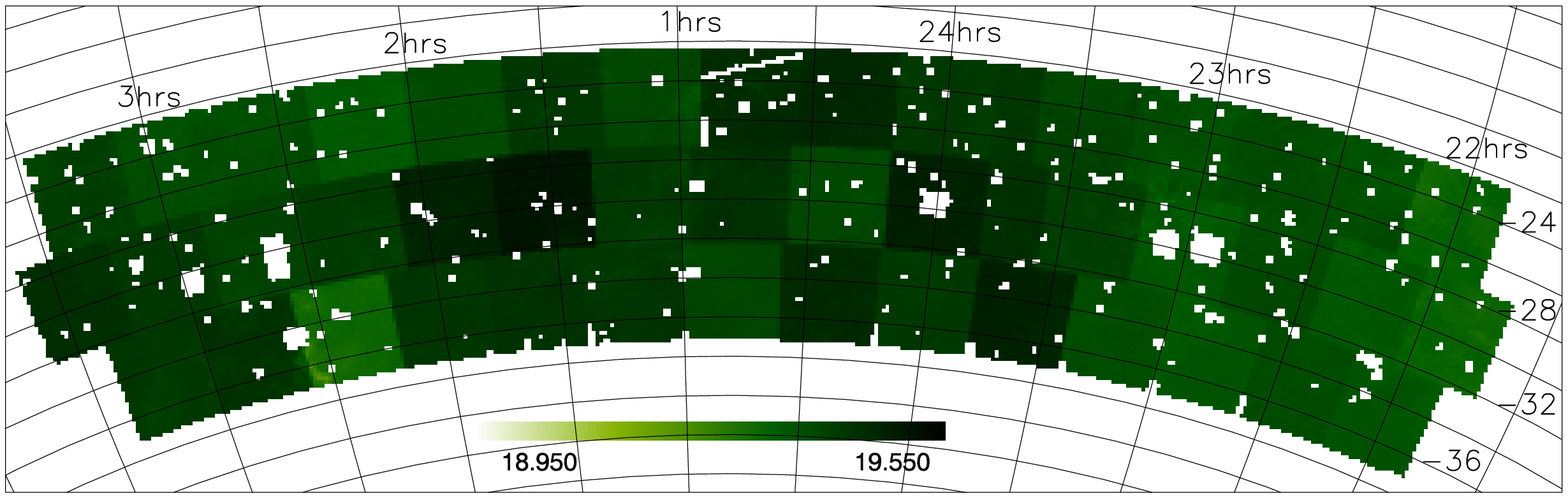}{1.0}
\caption{The magnitude limit masks for the NGP strip (top) and the SGP
strip (bottom), plotted in a zenithal equal area projection.}
\label{fig:maglimmask}
\end{figure*}

For accurate statistical analysis of the 2dFGRS it is essential to fully
understand the criteria that define its parent photometric catalogue and
also the spatial and magnitude-dependent completeness of the redshift
catalogue. For this purpose we have defined three maps or masks
characterizing this information as a function of position on the sky:

(1) The magnitude limit mask gives the extinction-corrected magnitude
limit of the survey at each position.

(2) The redshift completeness mask gives the fraction of measured
redshifts at each position.

(3) The magnitude completeness mask gives a parameter defining how the
redshift success rate depends on apparent magnitude.

Each mask has its own use, but for some analyses it is necessary to make
use of two or even all three masks. We now describe in more detail how
each one of these masks is defined and briefly outline some of their
uses.

\subsection{Magnitude limit mask}
\label{ssec:maglimmask}

Although the 2dFGRS sample was originally selected to have a uniform
extinction-corrected magnitude limit of \bj=19.45, in fact the survey
magnitude limit varies slightly with position on the sky. There are two
reasons for this. First, the photometric calibrations now available are
much more extensive than when the parent 2dFGRS catalogue was originally
defined. This has enabled us to recalibrate the whole 2dFGRS parent
catalogue (Maddox \etal\ 2001, in preparation), and results in new
zero-point offsets and linearity corrections for each of the UKST
photographic plates. Second, the extinction corrections have been
changed to use the final published version of the Schlegel \etal\ (1998)
extinction maps; the original extinction corrections came from a
preliminary version of those maps.

The magnitude limit mask is therefore defined by the change in the
photometric calibration of each UKST photographic plate and the change
in the dust extinction correction at each position on the sky. The
magnitude limit masks for the NGP and SGP strips are shown in
Figure~\ref{fig:maglimmask}; note that the mask also accounts for the
holes in the source catalogue around bright stars and plate flaws.

In the SGP, which is a subset of the APM galaxy survey (Maddox \etal\
1990a,b,c), the rms change in plate zero-point is only 0.03\,mag.
However, in the NGP region the original calibration was less accurate
and the change in zero-points have an rms of 0.08\,mag. The change in
the dust corrections are also less in the SGP, as the extinction is
generally lower in this region. In the SGP the rms magnitude change due
to improved dust corrections is 0.01\,mag while in the NGP it is
0.02\,mag.

\begin{figure}
\plotone{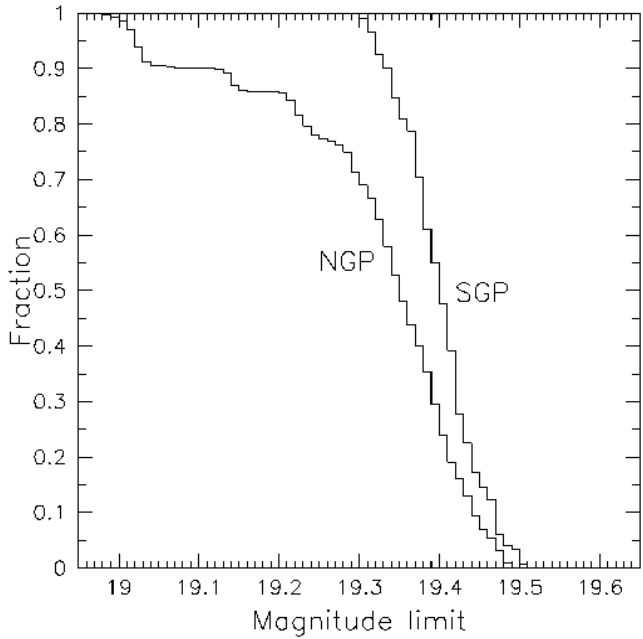}
\caption{The fraction of the sky in the NGP and SGP survey strips where
the survey limit is fainter than a given \bj\ magnitude. \vspace{30pt} }
\label{fig:maglimdist}
\end{figure}

The magnitude limit distribution over the NGP and SGP strips is shown in
Figure~\ref{fig:maglimdist}. In the SGP the median limiting magnitude is
\bj=19.40 with an rms about this value of 0.05\,mag; in the NGP the
median limiting magnitude is \bj=19.35 with an rms of 0.11\,mag.

For accurate statistical analysis of the 2dF survey the magnitude limits
defined by this mask should be used. It is always possible to analyse
the data with a fixed magnitude limit if one is prepared to omit both
the areas of the survey that have magnitude limits brighter than the
chosen limit and also all the galaxies in the remaining areas with
magnitudes fainter than the chosen limit.

\subsection{Simple redshift completeness mask}
\label{ssec:redmask}

\begin{figure*}
\plotfull{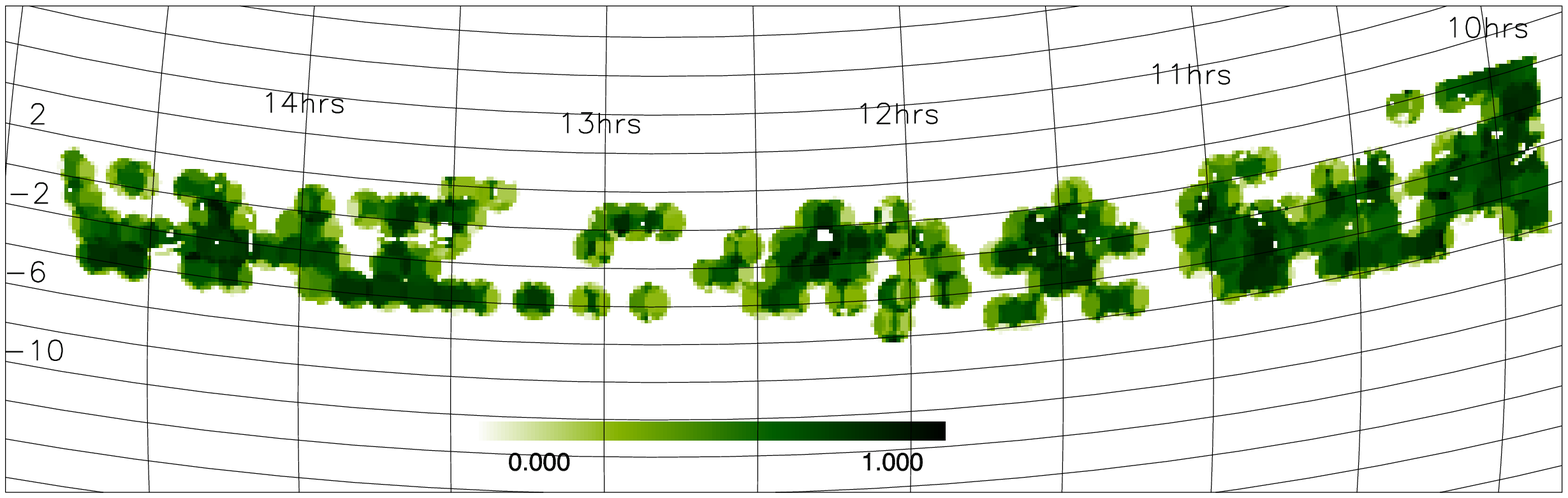}{1.0} \vspace{1pt} \\
\plotfull{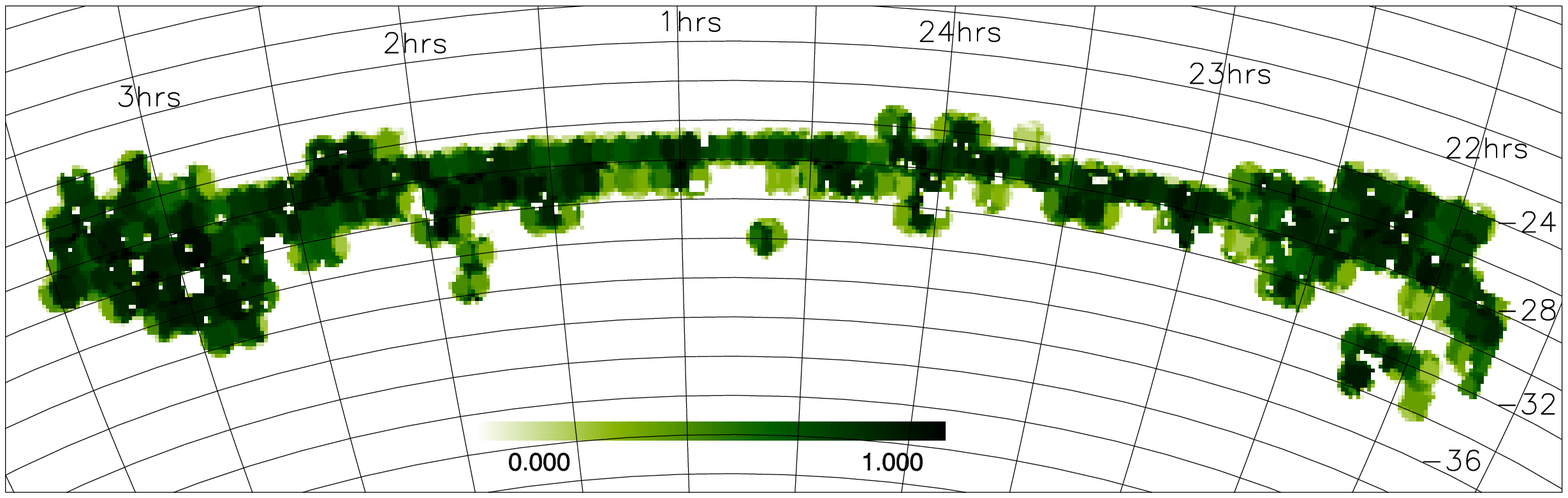}{1.0}
\caption{The redshift completeness, $R(\theta) \equiv N_z(\theta) /
N_{\rm p}(\theta)$ as a function of position for the 100k public release
dataset (2dFGRS data from October 1997 to January 2000). The top panel
is for the NGP, bottom panel for the SGP, plotted in a zenithal equal
area projection.}
\label{fig:zcompmask}
\end{figure*}

The best way to define a redshift completeness mask is to make use of
the geometry defined by the complete set of 2\degr\ fields that were
used to tile the survey region for spectroscopic observations. Each
region of the sky inside the survey boundary is covered by at least one
2\degr\ field, but more often by several overlapping fields. We define a
sector as the region delimited by a unique set of overlapping 2\degr\
fields. This is the most natural way of partitioning the sky, as it
takes account of the geometry imposed by the pattern of 2\degr\ fields
and the way in which the galaxies were targeted for spectroscopic
observation. Within each sector, $\theta$, we define the redshift
completeness, $R(\theta)$, as the ratio of the number of galaxies for
which redshifts have been obtained, $N_z(\theta)$, to the total number
of objects contained in the parent catalogue, $N_{\rm p}(\theta)$:
\begin{equation}
R(\theta) = N_z(\theta)/N_{\rm p}(\theta) .
\end{equation}
The redshift completeness of a given sector, $R(\theta)$, should be
clearly distinguished from the redshift completeness of a given field,
\CF, since multiple overlapping fields can contribute to a single
sector.

This simple redshift completeness mask, shown in
Figure~\ref{fig:zcompmask}, can be used to locate regions in which the
redshift completeness is high. It can also be used as a first step in
either applying weights to statistically correct for incompleteness or
to construct a random unclustered catalogue (for use in estimating
correlation functions) that have the same angular pattern of
incompleteness as the redshift sample. For this latter purpose one
should also take account of how the redshift completeness depends on
position within a sector as a result of constraints on fibre positioning
and other considerations. This is best done by using the parent
catalogue to derive weights for each galaxy with a measured redshift
(Norberg \etal\ 2001b, in preparation). Also, as discussed in the next
section, one should take account of how the redshift completeness
depends on apparent magnitude.

\subsection{Magnitude completeness mask}
\label{ssec:magmask}

The success rate of measuring redshifts is generally very high. Fields
for which the field completeness, \CF, is less than 70\% are
re-observed, while of the remainder just over 76\% have a completeness
greater than 90\%. As one approaches the magnitude limit of the survey
it becomes increasingly difficult to obtain good-quality spectra from
which reliable redshifts can be measured. Hence the success of measuring
redshifts (the redshift completeness) is a function of apparent
magnitude.

\begin{figure}
\plotone{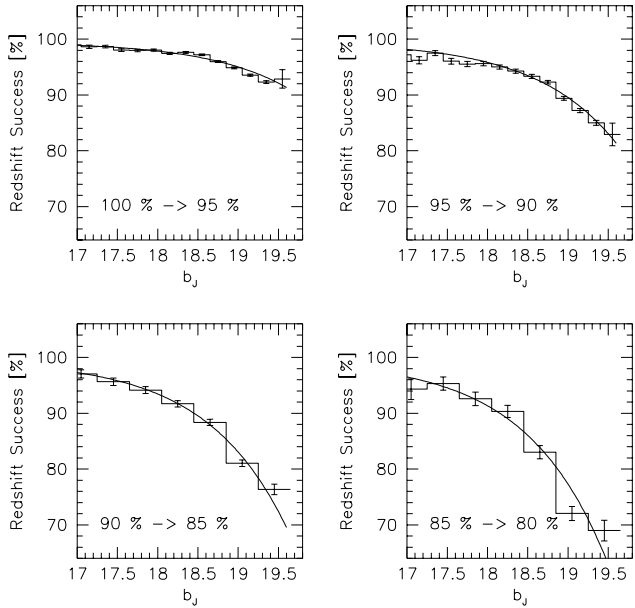} 
\caption{Redshift completeness as a function of apparent magnitude. The
different panels are for fields of different overall field completeness:
top left, fields with completeness in the range $95\%<\CF\le100\%$; top
right, fields with $90\%<\CF\le95\%$; bottom left, fields with
$85\%<\CF\le90\%$; and bottom right, fields with $80\%<\CF\le85\%$. Note
that over 76\% of the observed fields fall in the first two bins, with
$\CF>90\%$. The dashed curves are one-parameter model fits (see text).}
\label{fig:magcomplete}
\end{figure}

In Figure~\ref{fig:magcomplete} we show the redshift completeness as a
function of apparent magnitude for four different intervals of field
completeness, \CF. We see that in all cases the completeness is a
function of apparent magnitude, and that the magnitude at which the
completeness begins to drop is brightest in the fields with lowest \CF,
which are generally those taken in marginal observing conditions.
Overall, the magnitude-dependent incompleteness is well modelled by a
function of the form
\begin{equation}
c_z(m,\mu)=\gamma(1-\exp(m-\mu)) 
\label{eq:magcomp}
\end{equation}
with $\gamma=0.99$ and the parameter $\mu$ depending on \CF. For each
field we have chosen to fix $\mu$ by combining
equation~(\ref{eq:magcomp}) with a simple power-law model for the galaxy
number counts, $N(m) \propto \exp(\alpha m )$, so that \CF\ and $\mu$
are related by
\begin{equation}
\CF(\mu)=\frac{\int_{m_1}^{m_2}N(m)c_z(m,\mu)\,dm}{\int_{m_1}^{m_2}N(m)\,dm} ,
\label{eq:numcount}
\end{equation}
where $m_1$ and $m_2$ are the bright and faint magnitude limits. This
integral can be evaluated to give
\begin{eqnarray}
\CF(\mu)=\gamma\,(1-g(\alpha)\,\exp(m_1-\mu)) ,
\label{eq:numcount1}
\end{eqnarray}
where
\begin{eqnarray}
g(\alpha)=\frac{\alpha}{\alpha+1}
\frac{\exp\left[(\alpha+1)(m_2-m_1)\right]-1}{\exp\left[\alpha(m_2-m_1)\right]-1}
\end{eqnarray}
and we take $\alpha$=0.5 as appropriate for the galaxy number counts
around \bj=19. This equation can be inverted, yielding
\begin{equation}
\mu=\ln\left[\frac{ g(\alpha)\,\exp(m_1)}{(1-\CF/\gamma)}\right] ,
\label{eq:numcount2}
\end{equation}
which enables $\mu$ to be computed for each observed field.

Our goal is to define the value of $\mu$ characterising the
magnitude-dependent completeness for each position in the sky. Since the
2\degr\ fields overlap it again makes sense to define $\mu(\theta)$ for
each sector $\theta$. The value of $\mu(\theta)$ is defined by an
appropriately-weighted average of the $\mu$ values of the $N_{\rm
F}(\theta)$ overlapping observed fields comprising the sector.
Specifically we take
\begin{eqnarray}
\mu(\theta)=-\ln\left[\sum_{i=1}^{N_{\rm F}(\theta)}f_i\,\exp(-\mu_i)\right] ,
\label{eq:mutheta}
\end{eqnarray}
where $f_i$ is the fraction of the observed galaxies in this sector that
were targeted in field $i$.

Finally, it is straightforward to combine the magnitude completeness
given by $\mu(\theta)$ with the redshift completeness $R(\theta)$ to
define an estimate of redshift completeness that depends on both
position (sector) and magnitude:
\begin{eqnarray}
S(\theta,m)=\frac{N_{\rm p}(\theta)}{N_{\rm e}(\theta)}\,R(\theta)\,c_z(m,\mu(\theta)) .
\label{eq:comps}
\end{eqnarray}
In the first factor, $N_{\rm p}(\theta)$ is again the number of parent
catalogue galaxies in this sector while
\begin{eqnarray}
 N_{\rm e}(\theta)=\sum_{i=1}^{N_p(\theta)}c_z(m_i,\mu(\theta))
\end{eqnarray}
is an estimate of the number of galaxies one would expect to have
measured redshifts for given the value of $\mu$ that has been assigned
to this sector. For ease in calculating equation~(\ref{eq:comps}) we
tabulate $N_{\rm p}(\theta)/N_{\rm e}(\theta)$ as well as $\mu(\theta)$
and $R(\theta)$, using, for convenience, a pixellated mask rather than
the exact sectors.

\section{SURVEY DATABASE}
\label{sec:database}

The 2dFGRS database consists of three main components: (i)~a collection
of FITS files, one per object in the source catalogue, that contains
every piece of information about each object from the source catalogue,
spectroscopic observations and subsequent analysis; (ii)~an mSQL
database (Jepson \& Hughes 1998), that contains all of the parameters
for each object and allows complex searching and subsetting of the
survey objects and the retrieval of selected subsets of spectra and
images; and (iii)~a WWW interface, that provides a number of different
modes for querying the mSQL database and a variety of ways for returning
the results of such queries. The next three subsections describe these
three database components.

\subsection{Object FITS files}
\label{ssec:fits}

There are \totphot\ target objects in the FITS database, each with its
own FITS file. This number is much larger than the number of objects
surveyed, since it includes objects down to \bj=19.6 in the NGP and SGP
strips. Each object in the survey source catalogue has been given a
serial number ({\tt SEQNUM}), and the name of the FITS file for that
object is this serial number. The serial numbers for objects in the SGP
strip are 1--193550 and 416694--467214 for objects in the NGP strip
193551--332694 and 389714--416693, and for objects in the random fields
332695--389713.

Each FITS file has a primary part which contains all the source catalogue
data about the object (as FITS keywords) and a DSS postage stamp image
of the object. Spectra are appended as additional FITS extensions. Each
spectrum extension contains the spectrum of the object, the variance
(error) array for the object spectrum, the spectrum of the mean sky that
was subtracted from the object spectrum, and FITS keyword data giving
information about the spectroscopic observation and derived parameters
such as the redshift and spectral quality. Many targets contain multiple
spectrum extensions corresponding to multiple observations.

Table~\ref{tab:fits0} lists all the FITS keywords present in the primary
part (extension 0) of the FITS files; Table~\ref{tab:fits1} lists all
the keywords present in the spectrum extensions (1$\ldots${\tt
spectra}). The tables give the names of the keywords, example values,
and the keyword definitions.

\begin{table}
\centering
\caption{FITS keywords for the primary image (extension 0)}
\label{tab:fits0}
\begin{tabular}{lrl}
Keyword        &                  Example & Definition \vspace{6pt} \\
{\xx SIMPLE   }&{\xx                    T}&{\xx file does conform to FITS standard}\\
{\xx BITPIX   }&{\xx                   16}&{\xx number of bits per data pixel}\\
{\xx NAXIS    }&{\xx                    2}&{\xx number of data axes}\\
{\xx NAXIS1   }&{\xx                   49}&{\xx length of data axis 1}\\
{\xx NAXIS2   }&{\xx                   49}&{\xx length of data axis 2}\\
{\xx EXTEND   }&{\xx                    T}&{\xx FITS dataset may contain extensions}\\
{\xx BSCALE   }&{\xx               1.0000}&{\xx REAL = (FITS $\times$ BSCALE) + BZERO}\\
{\xx BZERO    }&{\xx               0.0000}&{\xx zeropoint of conversion to REAL}\\
{\xx SEQNUM   }&{\xx               100100}&{\xx Serial number : database primary key}\\
{\xx NAME     }&{\xx 'TGS469Z164'        }&{\xx 2dFGRS assigned name}\\
{\xx IMAGE    }&{\xx 'SKYCHART'          }&{\xx Existence of postage stamp image}\\
{\xx RA       }&{\xx         0.7943429758}&{\xx RA  (B1950) in radians :   3  2  3.00}\\
{\xx DEC      }&{\xx        -0.5475286941}&{\xx DEC (B1950) in radians : -31 22 15.9}\\
{\xx EQUINOX  }&{\xx              1950.00}&{\xx equinox of RA and DEC}\\
{\xx BJSEL    }&{\xx                18.93}&{\xx final \bj\ mag used in the object selection}\\
{\xx PROB     }&{\xx               2335.4}&{\xx psi star-galaxy classification parameter}\\
{\xx PARK     }&{\xx                0.910}&{\xx k   star-galaxy classification parameter}\\
{\xx PARMU    }&{\xx                0.187}&{\xx mu  star-galaxy classification parameter}\\
{\xx IGAL     }&{\xx                    1}&{\xx final classification flag (1 for a galaxy)}\\
{\xx JON      }&{\xx                   -1}&{\xx eyeball classification flag}\\
{\xx ORIENT   }&{\xx                 91.0}&{\xx orientation in degrees clockwise from E-W}\\
{\xx ECCENT   }&{\xx                0.270}&{\xx eccentricity}\\
{\xx AREA     }&{\xx                308.0}&{\xx isophotal area in pixels}\\
{\xx X\_BJ    }&{\xx               2918.7}&{\xx plate $x_{\bj}$ in 8 micron pixels}\\
{\xx Y\_BJ    }&{\xx               9123.1}&{\xx plate $y_{\bj}$ in 8 micron pixels}\\
{\xx DX       }&{\xx                 43.0}&{\xx corrected difference $100(x_{\bj}-x_R)$}\\
{\xx DY       }&{\xx                 49.0}&{\xx corrected difference $100(y_{\bj}-y_R)$}\\
{\xx BJG      }&{\xx                18.99}&{\xx \bj\ without extinction correction}\\
{\xx RMAG     }&{\xx                10.35}&{\xx unmatched APM `total' mag}\\
{\xx PMAG     }&{\xx                10.53}&{\xx unmatched raw APM profile integrated mag}\\
{\xx FMAG     }&{\xx                 8.72}&{\xx unmatched raw APM 2\arcsec\ `fibre' mag}\\
{\xx SMAG     }&{\xx                10.74}&{\xx unmatched raw stellar mag (from APMCAL)}\\
{\xx IFIELD   }&{\xx                  417}&{\xx UKST field }\\
{\xx IGFIELD  }&{\xx                 2007}&{\xx galaxy number in UKST field}\\
{\xx REGION   }&{\xx 'S417    '          }&{\xx GSSS region name}\\
{\xx OBJEQNX  }&{\xx              2000.00}&{\xx equinox of the plate reference frame}\\
{\xx OBJRA    }&{\xx         0.8034094522}&{\xx RA  (J2000) in radians :  03 04 07.673 }\\
{\xx OBJDEC   }&{\xx        -0.5441390223}&{\xx DEC (J2000) in radians : -31 10 36.73}\\
{\xx PLTSCALE }&{\xx              67.2000}&{\xx Plate scale in arcsec per mm}\\
{\xx XPIXELSZ }&{\xx           25.2844500}&{\xx X pixel size in microns}\\
{\xx YPIXELSZ }&{\xx           25.2844500}&{\xx Y pixel size in microns}\\
{\xx OBJPLTX  }&{\xx              7970.86}&{\xx object X on plate (pixels)}\\
{\xx OBJPLTY  }&{\xx              4148.11}&{\xx object Y on plate (pixels)}\\
{\xx DATAMAX  }&{\xx                14431}&{\xx Maximum data value}\\
{\xx DATAMIN  }&{\xx                 4011}&{\xx Minimum data value}\\
{\xx BJSELOLD }&{\xx                18.96}&{\xx original \bj\ mag used in the object selection}\\
{\xx BJG\_OLD }&{\xx                19.01}&{\xx original \bj\ without extinction correction}\\
{\xx END      }&{\xx                     }&{\xx End of FITS header}\\
\end{tabular}
\end{table}

\begin{table}
\centering
\caption{FITS keywords for the spectra (extensions 1$\ldots${\tt spectra})}
\label{tab:fits1}
\begin{tabular}{lrl} 
Keyword        &                     Example & Definition \\
{\xx XTENSION }&{\xx 'IMAGE   '             }&{\xx IMAGE extension}\\
{\xx BITPIX   }&{\xx                     -32}&{\xx number of bits per data pixel}\\
{\xx NAXIS    }&{\xx                       2}&{\xx number of data axes}\\
{\xx NAXIS1   }&{\xx                    1024}&{\xx length of data axis 1}\\
{\xx NAXIS2   }&{\xx                       3}&{\xx length of data axis 2}\\
{\xx PCOUNT   }&{\xx                       0}&{\xx required keyword; must = 0}\\
{\xx GCOUNT   }&{\xx                       1}&{\xx required keyword; must = 1}\\
{\xx CRVAL1   }&{\xx            5802.8979492}&{\xx coordinate value of axis 1}\\
{\xx CDELT1   }&{\xx            4.3103027344}&{\xx coordinate increment on axis 1}\\
{\xx CRPIX1   }&{\xx          512.0000000000}&{\xx reference pixel on axis 1}\\
{\xx CUNIT1   }&{\xx 'Angstroms'            }&{\xx units for axis 1}\\
{\xx EXTNAME  }&{\xx 'SPECTRUM'             }&{\xx 2dFGRS spectrum}\\
{\xx OBSNAME  }&{\xx 'TGS469Z164'           }&{\xx observed object name}\\
{\xx OBSRA    }&{\xx            0.7943429758}&{\xx observed RA  (B1950) in radians}\\
{\xx OBSDEC   }&{\xx           -0.5475286940}&{\xx observed DEC (B1950) in radians}\\
{\xx MATCH\_DR}&{\xx                  0.0000}&{\xx position match error in arcsec}\\
{\xx Z        }&{\xx                0.178876}&{\xx raw measured redshift}\\
{\xx Z\_HELIO }&{\xx                0.178860}&{\xx heliocentric redshift}\\
{\xx QUALITY  }&{\xx                       5}&{\xx redshift measurement quality}\\
{\xx ABEMMA   }&{\xx                       1}&{\xx redshift type: abs=1,emi=2,man=3}\\
{\xx NMBEST   }&{\xx                       0}&{\xx \# emission lines for emission z}\\
{\xx NGOOD    }&{\xx                       0}&{\xx number of good emission lines}\\
{\xx Z\_EMI   }&{\xx                 -9.9990}&{\xx emission redshift}\\
{\xx Q\_Z\_EMI}&{\xx                       0}&{\xx emission redshift quality}\\
{\xx KBESTR   }&{\xx                       2}&{\xx cross-correlation template}\\
{\xx R\_CRCOR }&{\xx                 15.5600}&{\xx cross-correlation peak}\\
{\xx Z\_ABS   }&{\xx                  0.1789}&{\xx cross-correlation redshift}\\
{\xx Q\_Z\_ABS}&{\xx                       3}&{\xx cross-correlation quality}\\
{\xx Q\_FINAL }&{\xx                       3}&{\xx suggested quality for redshift}\\
{\xx IALTER   }&{\xx                       0}&{\xx IALTER=1 if automatic z altered}\\
{\xx Z\_COMM  }&{\xx '        '             }&{\xx observer's comment}\\
{\xx THPUT    }&{\xx                 0.96613}&{\xx fibre throughput}\\
{\xx SPFILE   }&{\xx 'sgp469\_991104\_1z.fits'}&{\xx 2dF reduced data file}\\
{\xx PLATE    }&{\xx                       1}&{\xx 2dF plate number}\\
{\xx PIVOT    }&{\xx                     302}&{\xx 2dF pivot number}\\
{\xx FIBRE    }&{\xx                      58}&{\xx 2dF fibre number}\\
{\xx OBSRUN   }&{\xx '99OCT   '             }&{\xx observation run}\\
{\xx GRS\_DATE}&{\xx '991104  '             }&{\xx 2dF YYMMDD observed date}\\
{\xx UTDATE   }&{\xx '1999:11:04'           }&{\xx UT date of observation}\\
{\xx SPECTID  }&{\xx 'A       '             }&{\xx 2dF spectrograph ID}\\
{\xx GRATID   }&{\xx '300B    '             }&{\xx 2dF grating ID}\\
{\xx GRATLPMM }&{\xx                     300}&{\xx 2dF grating line per mm}\\
{\xx GRATBLAZ }&{\xx 'COLLIMATOR'           }&{\xx 2dF grating blaze direction}\\
{\xx GRATANGL }&{\xx                25.30000}&{\xx 2dF grating angle}\\
{\xx LAMBDAC  }&{\xx                5782.700}&{\xx central wavelength}\\
{\xx CCD      }&{\xx 'TEKTRONIX\_5'         }&{\xx CCD ID}\\
{\xx CCDGAIN  }&{\xx                   2.790}&{\xx CCD inverse gain (e per ADU)}\\
{\xx CCDNOISE }&{\xx                   5.200}&{\xx CCD read noise (electrons)}\\
{\xx OBJX     }&{\xx                  196833}&{\xx 2dF object X position}\\
{\xx OBJY     }&{\xx                   10401}&{\xx 2dF object Y position}\\
{\xx OBJXERR  }&{\xx                       6}&{\xx 2dF object X position error}\\
{\xx OBJYERR  }&{\xx                      14}&{\xx 2dF object Y position error}\\
{\xx OBJMAG   }&{\xx                   18.96}&{\xx 2dF object magnitude}\\
{\xx THETA    }&{\xx                   4.526}&{\xx 2dF fibre angle}\\
{\xx PTRTYPE  }&{\xx 'P       '             }&{\xx 2dF ptrtype}\\
{\xx PID      }&{\xx                       0}&{\xx 2dF pid}\\
{\xx OBSFLD   }&{\xx 'sgp469  '             }&{\xx 2dF observed field number}\\
{\xx NCOMB    }&{\xx                       3}&{\xx number of frames combined}\\
{\xx REFRUN   }&{\xx                      31}&{\xx AAT run number of reference}\\
{\xx UTSTART  }&{\xx '16:37:59.48'          }&{\xx UT start of reference exposure}\\
{\xx UTEND    }&{\xx '16:57:59'             }&{\xx UT end of reference exposure}\\
{\xx REFEXP   }&{\xx                  1200.0}&{\xx reference run exposure (secs)}\\
{\xx REFHASTA }&{\xx '36.07264'             }&{\xx HA start of reference exposure}\\
{\xx REFHAEND }&{\xx '41.08119'             }&{\xx HA end of reference exposure}\\
{\xx ETA\_TYPE}&{\xx          -2.5934501E+00}&{\xx eta spectral type parameter}\\
{\xx SNR      }&{\xx           2.0299999E+01}&{\xx median S/N per pixel}\\
\end{tabular}
\end{table}

\subsection{mSQL parameter database}
\label{ssec:database}

The mSQL database can be thought of as a table. The rows of the table
are labelled by the unique object serial number ({\tt SEQNUM}) and the
extension number ({\tt extnum}), with multiple rows for each target
object corresponding to each extension in the object's FITS file. The
columns of the table correspond to the object parameters, and are
labelled by the name of the corresponding keyword. The object serial
numbers ({\tt SEQNUM}) provide the primary database key, but the objects
are also indexed by their unique survey name ({\tt NAME}), which has the
format {\tt TGhfffZnnn}, where {\tt h} is the hemisphere (N for the NGP
strip and S for the SGP strip and random fields), {\tt fff} is the
number of the primary field to which the object is assigned and {\tt
nnn} is the number of the galaxy within that field. Note that the
observed name of the object (parameter {\tt OBSNAME} in each spectrum
extension) is the same as {\tt NAME} except that: (i)~if the field in
which the object is observed (given by {\tt OBSFLD}) is an overlapping
field rather than its primary field (given by {\tt fff}), then the first
character of the name is changed from {\tt T} to {\tt X}; and (ii)~if
the object has been flagged as a possible merger, then the second
character of the name is changed from {\tt G} to {\tt M}.

The first row for each object ({\tt extnum}=0) contains the source
catalogue data and the basic spectroscopic information from the best
spectrum of that object. The keywords for that row are the FITS
parameters for the primary image (Table~\ref{tab:fits0}) plus all the
additional keywords listed in Table~\ref{tab:msql}. The best spectrum is
the one with the highest redshift quality parameter; if there is more
than one spectrum of the same quality, then the most recent of these
spectra is used. Subsequent rows for the same object ({\tt
extnum}=1$\ldots${\tt spectra}, where {\tt spectra} is the number of
spectra obtained for that object) contain the FITS parameters pertaining
to each spectroscopic observation (Table~\ref{tab:fits1}) plus the
additional keywords in section~(i) of Table~\ref{tab:msql}. If there is
no spectrum for the object then {\tt spectra}=0 and only the row
corresponding to {\tt extnum}=0 will exist. Note that some information
is duplicated between rows and that not all parameters are defined for
all rows; undefined parameters return a {\tt NULL} value.

\begin{table}
\centering
\caption{Additional mSQL database keywords}
\label{tab:msql}
\begin{tabular}{lrl}
\multicolumn{3}{l}{(i) Keywords in all extensions} \vspace{6pt} \\
Keyword       &            Example & Definition \vspace{6pt} \\
{\xx serial  }&{\xx         100100}&{\xx 2dFGRS serial number}\\
{\xx name    }&{\xx     TGS469Z164}&{\xx 2dFGRS name}\\
{\xx UKST    }&{\xx            417}&{\xx UKST sky survey field number}\\
{\xx spectra }&{\xx              1}&{\xx number of spectra for this object}\\
{\xx extnum  }&{\xx              1}&{\xx extension number}\\
{\xx obsrun  }&{\xx          99OCT}&{\xx observing run year and month}\\
{\xx TDFgg   }&{\xx           -469}&{\xx 2dFGRS field number (+NGP, -SGP)}\\
{\xx pivot   }&{\xx            302}&{\xx 2dF pivot}\\
{\xx plate   }&{\xx              1}&{\xx 2dF plate}\\
{\xx fiber   }&{\xx             58}&{\xx 2dF fibre}\\
{\xx z       }&{\xx       0.178876}&{\xx observed redshift}\\
{\xx z\_helio}&{\xx       0.178860}&{\xx heliocentric redshift}\\
{\xx abemma  }&{\xx              1}&{\xx redshift type (abs=1,emi=2,man=3)}\\
{\xx quality }&{\xx              5}&{\xx redshift quality parameter}\\
              &                    & \\
\multicolumn{3}{l}{(i) Keywords in extension 0 only} \vspace{6pt} \\
Keyword       &            Example & Definition \vspace{6pt} \\
{\xx alpha   }&{\xx   0.7943429758}&{\xx RA  (B1950) in radians}\\
{\xx delta   }&{\xx  -0.5475286941}&{\xx DEC (B1950) in radians}\\
{\xx ra      }&{\xx     3  2  3.00}&{\xx RA  (B1950) in HH MM SS.SS}\\
{\xx dec     }&{\xx    -31 22 15.9}&{\xx DEC (B1950) in DD MM SS.S }\\
{\xx ra2000  }&{\xx    03 04 07.68}&{\xx RA  (J2000) in HH MM SS.SS}\\
{\xx dec2000 }&{\xx    -31 10 36.8}&{\xx DEC (J2000) in DD MM SS.S }\\
{\xx l2      }&{\xx 228.9258834424}&{\xx Galactic longitude}\\
{\xx b2      }&{\xx -60.8572447739}&{\xx Galactic latitude}\\
\end{tabular}
\end{table}

Searches of the database use the mSQL query format (Jepson \& Hughes
1998), which has the basic format
\begin{mSQL}
SELECT list\_of\_parameters FROM database\_name WHERE list\_of\_conditions
\end{mSQL}
and {\tt list\_of\_conditions} is a series of equalities and
inequalities linked by Boolean relations. 

An example is
\begin{mSQL}
SELECT name, extnum, ra, dec, BJSEL, Z, QUALITY, z, quality FROM public
WHERE name='TGS469Z164'
\end{mSQL}
which selects the listed parameters for the object with 2dFGRS name {\tt
TGS469Z164} (note the single quotes around the character string) from
both the summary row ({\tt extnum}=0) and for each spectrum ({\tt
extnum}=1$\ldots${\tt spectra}). Note that the parameters with the same
name in lower case and upper case are distinct: the former are generally
from {\tt extnum}=0, the latter from {\tt extnum}$>$0 (parameters are
returned as {\tt NULL} in rows where they are not defined). 

An example with a more complex list of conditions is
\begin{mSQL}
SELECT name FROM public WHERE extnum=0 AND ((BJSEL$<$15.5 AND
quality$>$=3) OR quality$>$4) 
\end{mSQL}
which lists just the names of the objects which are either brighter than
\bj=15.5 with redshift quality at least 3, or have quality greater than
4, or both; the search is restricted just to the summary row by
requiring {\tt extnum}=0. 

Simple searches on the two indexed parameters, {\tt serial} and {\tt
name}), are quick---e.g.\ {\tt WHERE serial=69656} or {\tt WHERE
name='TGS203Z081'}; more complex searches take about 5 minutes. Further
information about the mSQL database software and its structured query
language is given in Yarger \etal\ (1999) and on the WWW at
http://www.hughes.com.au.

\subsection{WWW interface}
\label{ssec:www}

The 2dFGRS mSQL database can be searched via the WWW interface in a
number of ways: (i)~search via a standard mSQL query as described
above---this is the most general method; (ii)~perform a standard mSQL
query restricted to a list of named objects; (iii)~perform a standard
mSQL query restricted to objects within a specified radius of a given
position on the sky; (iv)~match objects to a supplied catalogue of
positions. 

The results of a query can be returned in several forms: as an HTML
table, as an ASCII table, as a gzipped ASCII file, as an email giving
the URL of a gzipped ASCII file created with a background job, or as a
tar file containing the FITS files for the selected objects. The HTML
table has buttons allowing you to select a particular object and view
the postage stamp DSS image and all the observed spectra for the object;
if the spectra have measured redshifts, then the positions of prominent
spectral features are indicated at the redshift associated with each
individual spectrum. The FITS files for the objects in the HTML table
can also be bundled up into a compressed tar file which can then be down
loaded by anonymous ftp.

Full instructions for accessing the database are given on the survey
website at http://www.mso.anu.edu.au/2dFGRS and at mirror sites given
there.

\section{RESULTS}
\label{sec:results}

\begin{figure}
\plotone{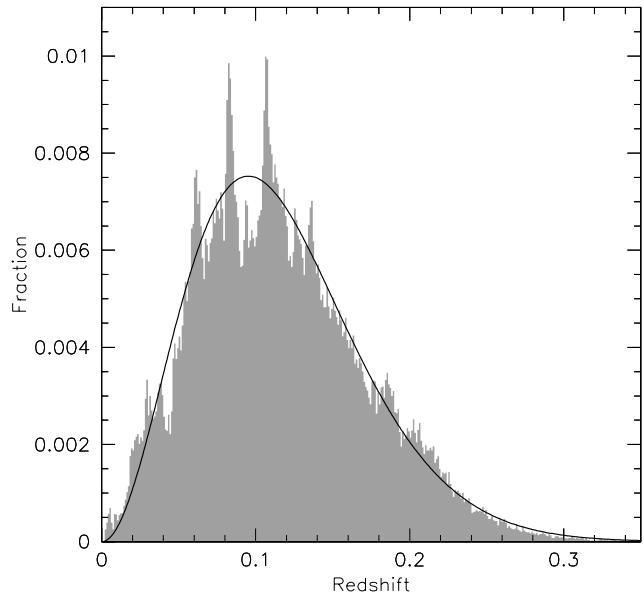} 
\caption{The redshift distribution of the 2dFGRS (histogram) and a
simple smooth analytic approximation (curve).}
\label{fig:2dFnz}
\end{figure}

The redshift distribution of the galaxies in the survey is shown in
Figure~\ref{fig:2dFnz}. Also shown is a simple analytic approximation to
the redshift distribution, following Efstathiou \& Moody (2001), with
the form
\begin{equation}
dN \propto z^2 \exp \left[ - \left( 
   \frac{1.36z}{\bar{z}} \right)^{1.55} \right]\,{\rm d}z
\end{equation}
where $\bar{z}$=0.11 is the median redshift of the survey.

\begin{figure*}
\plotrot{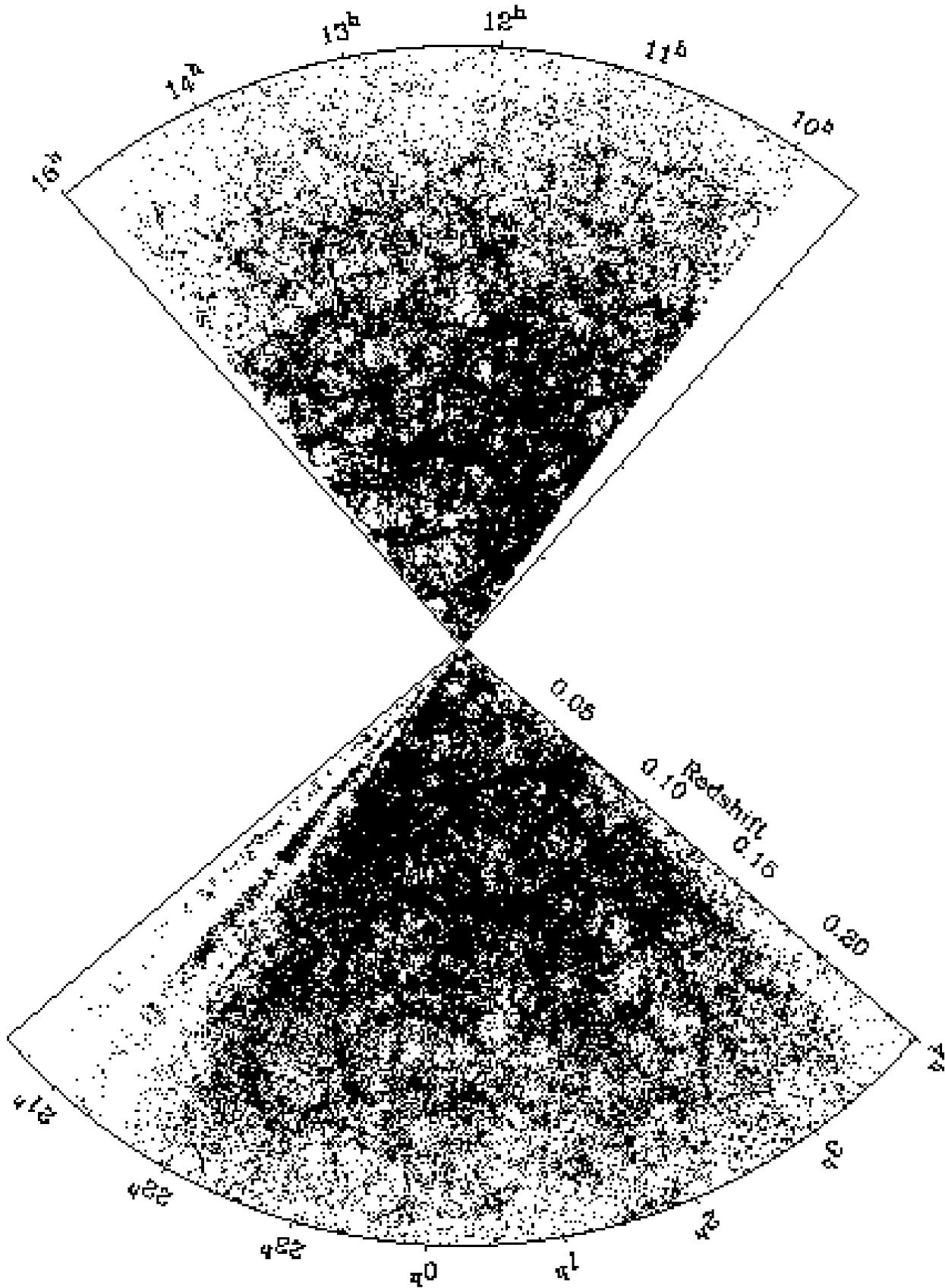}{0.95}{180}
\caption{The projected distribution of the galaxies in the NGP (top) and
SGP (bottom) strips, as a function of redshift and R.A.; the variations
in the galaxy density with R.A. are due to variations in the effective
widths of the strips.}
\label{fig:2dFzcone}
\end{figure*}

The spatial distribution of the galaxies in the survey strips is shown
in Figure~\ref{fig:2dFzcone}. This figure is the projection of the full
width of the strips (10\degr\ in the NGP and 15\degr\ in the SGP). As
not all fields have yet been observed, there are variations in the
effective thicknesses of the strips in the directions of the missing
fields that produce corresponding variations in the galaxy density. The
clustering of galaxies revealed in this figure has been investigated by
Peacock \etal\ (2001), Percival \etal\ (2001) and Norberg \etal\ (2001a).

The 2dFGRS has provided the first clear detection of the redshift-space
clustering anisotropy on large scales that is a key prediction of the
gravitational instability paradigm for the growth of structure in the
universe (Peacock \etal\ 2001). Measurements of this distortion yield a
precise estimate of the parameter $\beta = \Omega^{0.6}/b = 0.43 \pm
0.07$, where $\Omega$ is the total mass density of the universe and $b$
is the linear bias of the galaxies with respect to the mass. Combined
with recent measurements of the anisotropies in the cosmic microwave
background (Jaffe \etal\ 2001), this result favours a low-density
universe, with $\Omega$$\approx$0.3.

The power spectrum of the galaxy distribution has been determined from
the survey using a direct FFT-based technique (Percival \etal\ 2001).
Over the range in wavenumber 0.02$<$$k$$<$0.15\invMpc, the shape of the
observed power spectrum will be close to that of the linear density
perturbations convolved with the window function of the survey. Fitting
convolved power spectrum models constrains the shape parameter $\Gamma =
\Omega h$ to be $0.20 \pm 0.03$ (68\% confidence level) and shows that
models containing baryon oscillations are preferred over models without
baryons at the 95\% confidence level. This is the first detection of
baryon oscillations in the galaxy distribution, and yields an estimate
for the baryon fraction $\Omega_b/\Omega_m = 0.15 \pm 0.07$, assuming
scale-invariant primordial fluctuations.

The size of the 2dFGRS sample has allowed investigation of the variation
in the strength of galaxy clustering with luminosity, using the
projected two-point correlation function of galaxies in a series of
volume-limited samples (Norberg \etal\ 2001a). The clustering of $L^*$
($M_{b_J}^* - 5\log h = -19.7$) galaxies in real space is well fitted by
a power-law relation with exponent $\gamma = 1.71 \pm 0.06$ and a
correlation length $r_0^* = (4.9\pm0.3)\Mpc$. The exponent shows very
little variation for galaxy samples with luminosities differing by a
factor of 40. The clustering amplitude, however, increases with
luminosity: slowly for galaxies fainter than $M^*$, but more strongly at
brighter absolute magnitudes. This dependence of the correlation length
on luminosity is in good agreement with the predictions of the
hierarchical galaxy formation models of Benson \etal\ (2001). In terms
of the bias parameter, the relation is well-represented by $\delta b/b^*
= 0.15 \delta L/L^*$, where $\delta b = b-b^*$, $\delta L = L-L^*$ and
the relative bias is given by $b/b^* = (r_0/r_0^*)^{\gamma/2}$.

The 2dFGRS has also been used to characterise the internal properties of
the galaxy population. Folkes \etal\ (1999) and Madgwick \etal\ (2001a)
have determined the luminosity function for galaxies, both overall and
as a function of spectral type. Principal Component Analysis has been
applied to the 2dFGRS spectra, and a linear combination of the first two
principal components, $\eta$, has been used to parametrise the spectral
type. Going from early types to late types, the luminosity functions
appear to exhibit a systematic decrease in the characteristic luminosity
(from $M_{b_J}^* - 5\log h = -19.6$ to $-19.0$) and a steepening of the
faint-end slope (from $\alpha = -0.52$ to $-1.43$). However there is
also evidence that, at the precision afforded by the 2dFGRS sample, the
standard Schechter fitting function is no longer an adequate
representation of the luminosity function.

\begin{figure*}
\plotrot{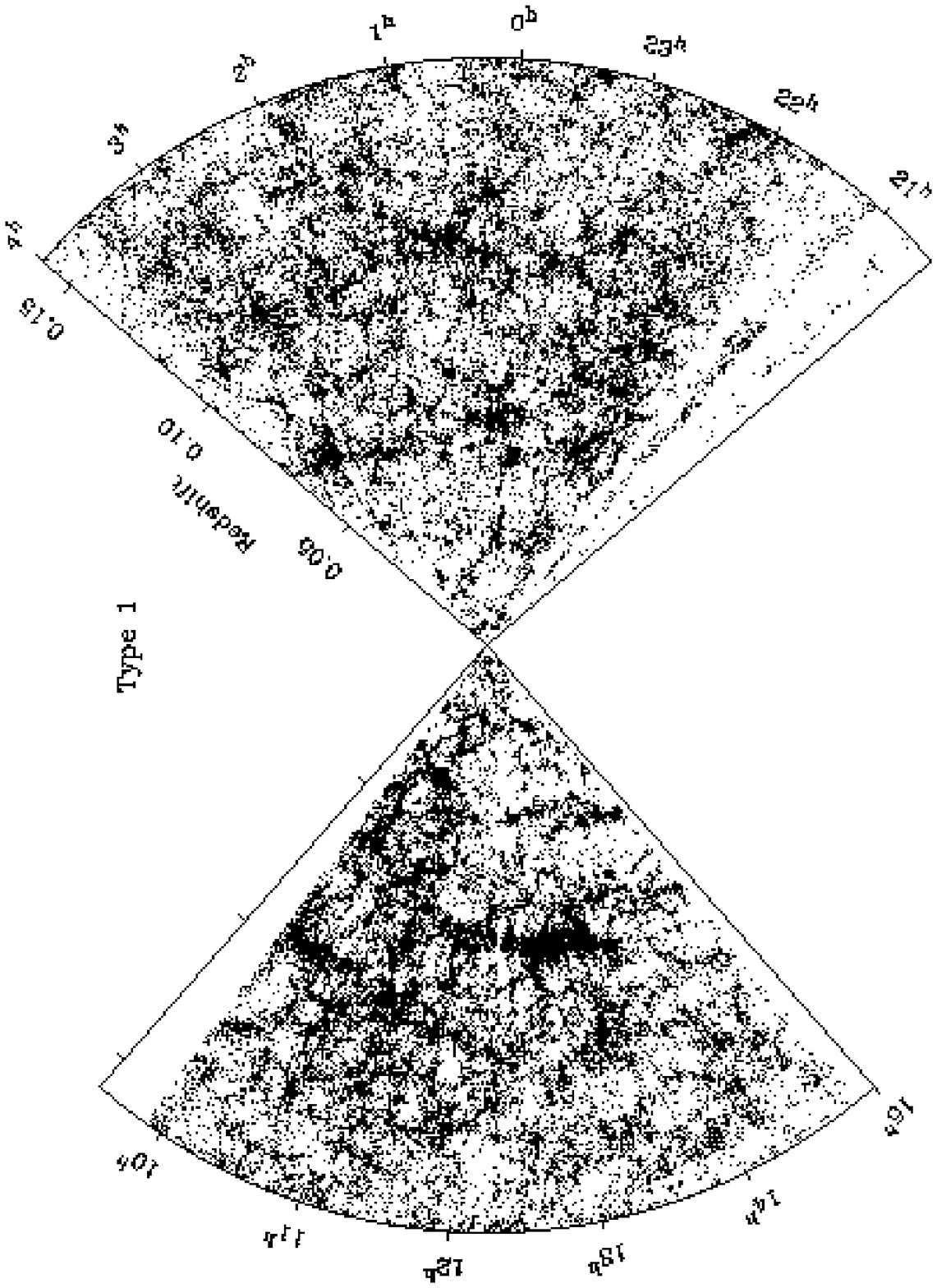}{0.9}{270} ~ \\ ~ \\
\plotrot{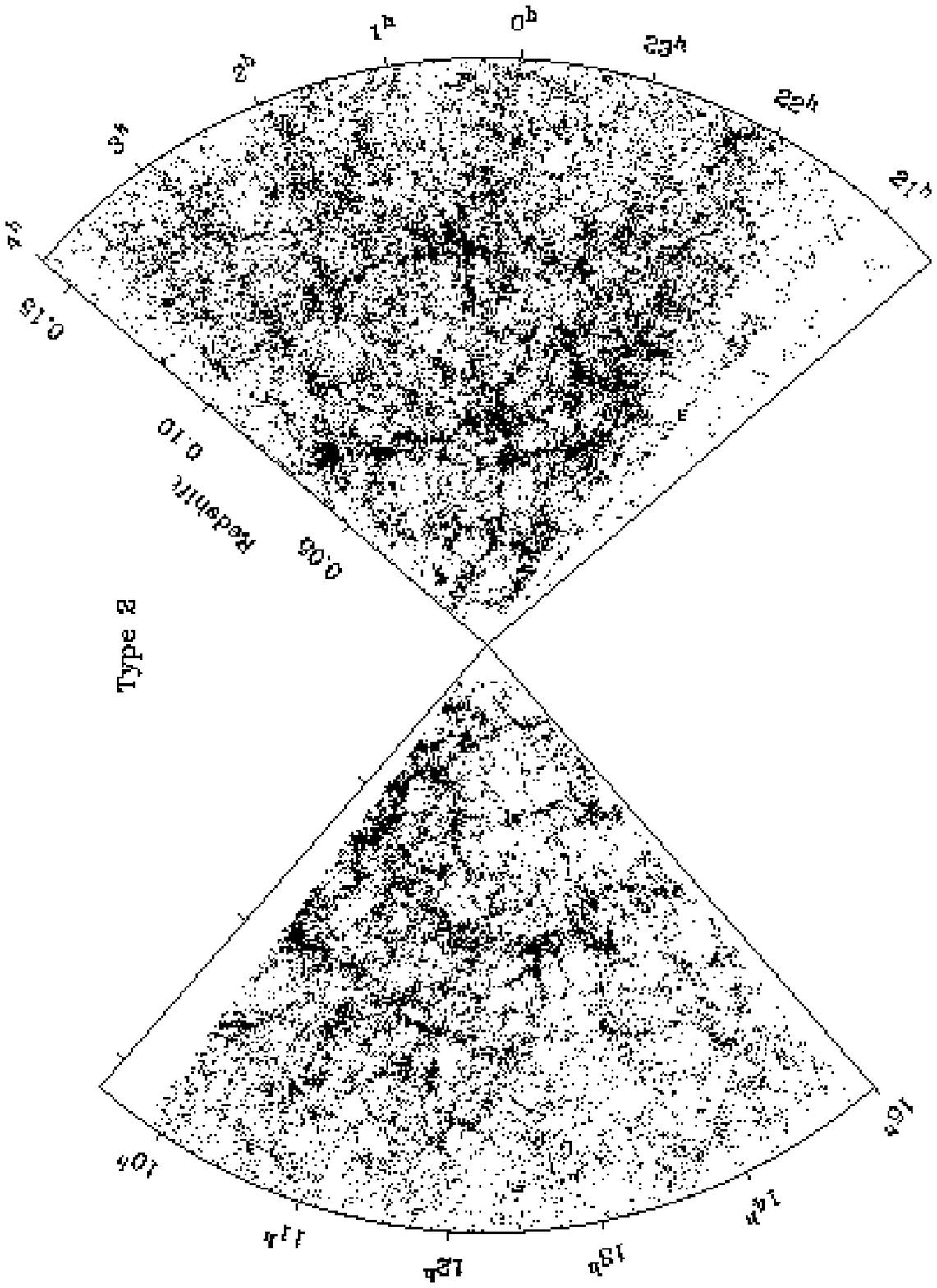}{0.9}{270}
\caption{Redshift slices for different spectral types: type~1
corresponds to E/S0, type~2 to Sa/Sb, type~3 to Sc/Sd and type~4 to
Irr.}
\label{fig:etacones}
\end{figure*}

Figure~\ref{fig:etacones} shows the redshift slices for each of the four
spectral types defined in terms of the $\eta$ parameter by Madgwick
\etal\ (2001a): type~1 are the earliest types and type~4 the latest
types. The association of earlier types with clusters and local density
enhancements is apparent, as is the tendency for later types to be
associated with lower-density regions. A quantitative study of the
clustering of different spectral types will be the subject of a
subsequent paper.

\addtocounter{figure}{-1}
\begin{figure*}
\plotrot{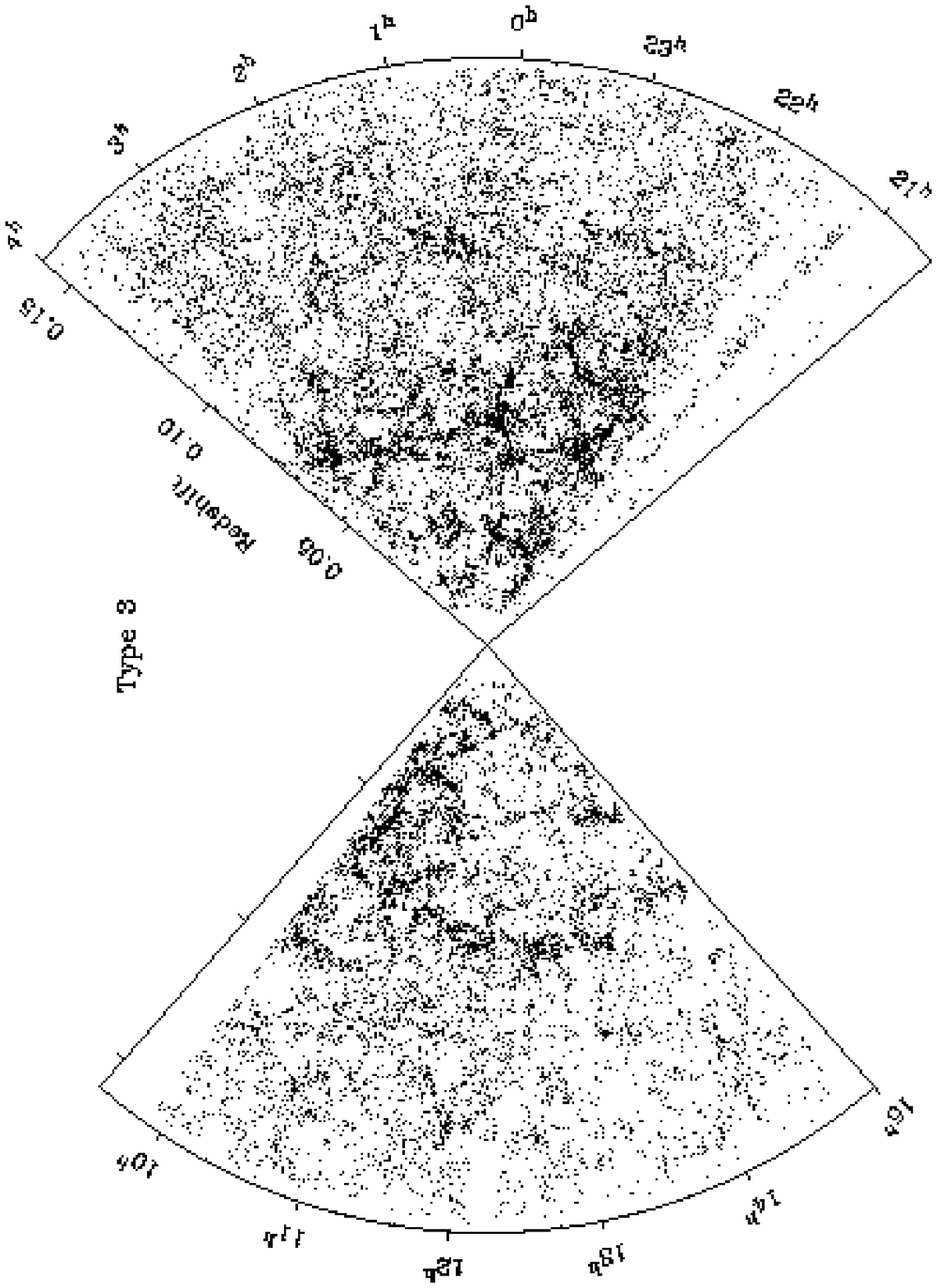}{0.9}{270} ~ \\ ~ \\
\plotrot{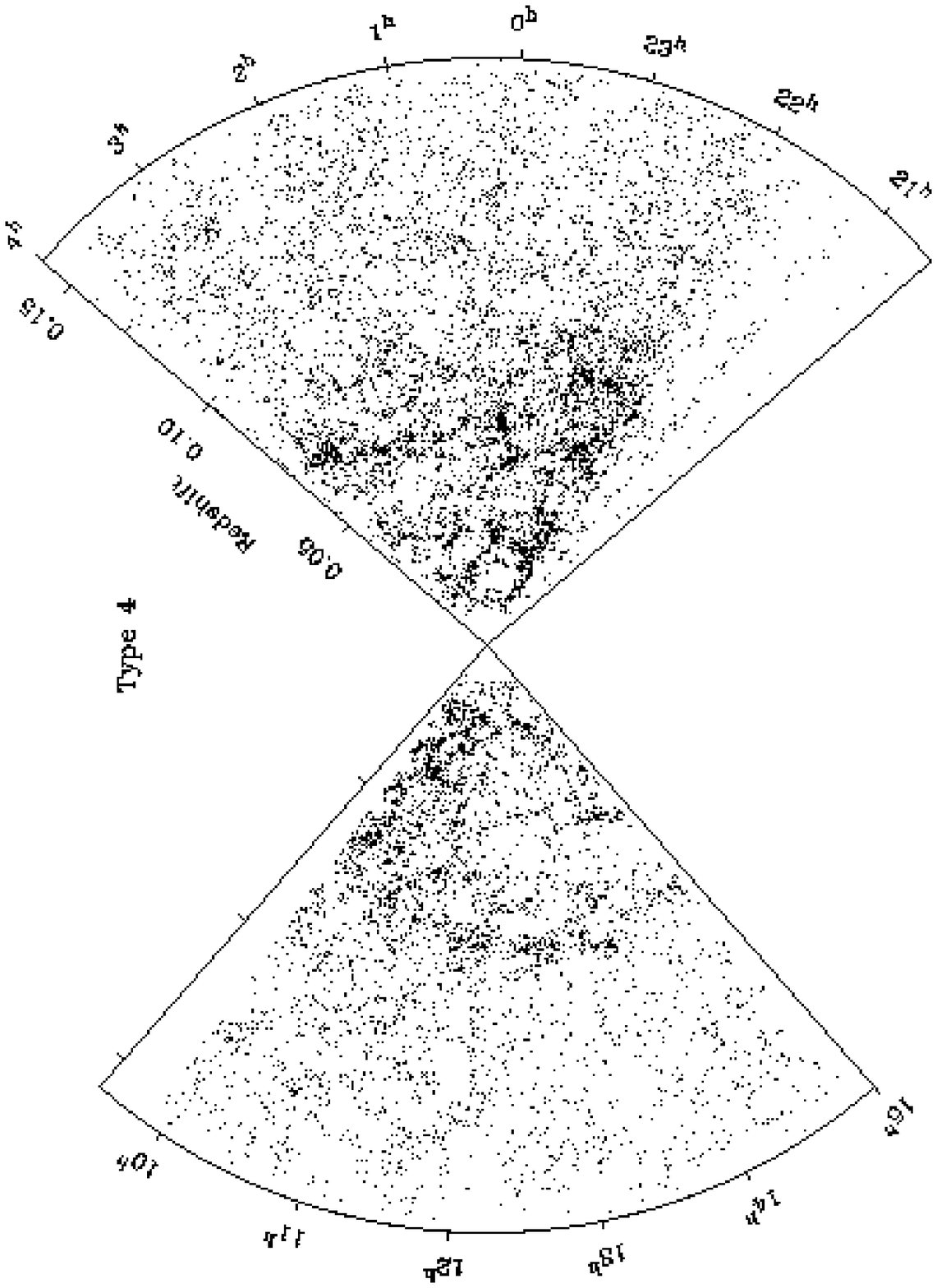}{0.9}{270}
\caption{(ctd)}
\end{figure*}

The large 2dFGRS sample also allows the generalisation of the luminosity
function into the bivariate brightness distribution (BBD) over both
luminosity and surface brightness (Cross \etal\ 2001). The BBD derived
from the 2dFGRS shows a strong surface brightness--luminosity relation,
$M_{b_J} \propto (2.4\pm^{1.5}_{0.5}) \mu_e$. The luminosity density is
dominated by normal giant galaxies and the peak of the BBD lies away
from the survey selection boundaries, implying that the 2dFGRS is
complete and that luminous low surface brightness galaxies are rare. By
integrating over the BBD, the local luminosity density is estimated to
be $j_B = (2.5\pm0.2)\times10^8\,h\,L_{\odot}\,{\rm Mpc}^{-3}$.

Cole \etal\ (2001) have combined the 2dFGRS with the 2MASS extended
source catalogue to produce an infrared-selected sample of over 17\,000
galaxies with redshifts. This sample has been used to determine the $J$
and $K_S$-band galaxy luminosity functions. The luminosity functions (in
2MASS Kron magnitudes) are fairly well fit by Schechter functions: in
$J$ with $M_J^* - 5\log h = -22.36\pm0.02$, $\alpha_J = -0.93\pm0.04$,
$\Phi_J^* = 0.0104\pm0.0016\invcubicMpc$, and in $K_S$ by $M_K^* - 5\log
h = -23.44\pm0.03$, $\alpha_K = -0.96\pm0.05$, $\Phi_K^* =
0.0108\pm0.0016\invcubicMpc$. From the distributions of $B$$-$$K$ and
$J$$-$$K$ colours with absolute magnitude and models of the stellar
populations, the galaxy stellar-mass function can be estimated.
Integrated over all galaxy masses, it yields a total mass fraction in
stars (in units of the critical density) of $\Omega_{\rm stars}h =
(1.6\pm0.2)\times10^{-3}$ for a Kennicutt IMF, and $\Omega_{\rm stars}h
= (2.9\pm0.4)\times10^{-3}$ for a Salpeter IMF. These values are
consistent with estimates based on the time integral of the observed
star formation history of the universe only if dust extinction
corrections at high redshift are modest.

The 2dFGRS can add significant value to all-sky imaging surveys at all
wavelengths, as in the studies of the radio galaxy population combining
redshifts and optical spectra from the 2dFGRS with the NVSS radio survey
(Sadler \etal\ 2001) and the FIRST radio survey (Magliocchetti \etal\
2001). Using 20\% of the full 2dFGRS area, Sadler \etal\ (2001) find 757
optical counterparts for NVSS sources---the largest and most homogeneous
set of radio-source spectra to date. These sources range from $z$=0.005
to $z$=0.438, and are a mixture of active galaxies (60\%) and
star-forming galaxies (40\%). The local radio luminosity function at 1.4
GHz is determined for both active and star-forming galaxies, and yields
an estimate for the local star-formation density of
$(2.8\pm0.5)\times10^{-3}\,M_{\odot}\,{\rm yr}^{-1}\invcubicMpc$.

\section{CONCLUSIONS}
\label{sec:conclusions}

The 2dF Galaxy Redshift Survey (2dFGRS) is designed to measure redshifts
for approximately 250\,000 galaxies. The survey uses the 2dF multi-fibre
spectrograph on the Anglo-Australian Telescope, which is capable of
observing up to 400 objects simultaneously over a 2\degr\ diameter field
of view. The source catalogue for the survey is a revised and extended
version of the APM galaxy catalogue, which is based on Automated Plate
Measuring machine (APM) scans of 390 plates from the UK Schmidt
Telescope (UKST) Southern Sky Survey. The target galaxies have
extinction-corrected magnitudes brighter than \bj=19.45, with
extinctions derived from the maps of Schlegel \etal\ (1998).

The main survey regions are two contiguous declination strips, one in
the southern Galactic hemisphere (the SGP strip) covering
80$\degr$$\times$15$\degr$ centred close to the SGP, and the other in
the northern Galactic hemisphere (the NGP strip) covering
75$\degr$$\times$10$\degr$. In addition, there are 99 random fields are
spread uniformly over the entire region of the APM catalogue in the
southern Galactic hemisphere outside the SGP strip. In total the survey
covers some 2000\,deg$^2$ and has a median depth of $\bar{z}$=0.11. Out
to the effective limit of the survey at $z$$\approx$0.3, the strips
contain a volume of 1.2$\times$10$^8$\cubicMpc; the volume sparsely
sampled including the random fields is two to three times larger. An
adaptive tiling algorithm is used to give a highly uniform sampling rate
of 93\% over the whole survey region.

Spectra are obtained over the wavelength range 3600--8000\AA\ at
two-pixel resolution of 9.0\AA. The median $S/N$ is 13\perpix\ over
4000--7500\AA. Redshifts are measured both by cross-correlation with a
range of template spectra and by fitting strong spectral features. All
redshift identifications are visually checked and assigned a quality
parameter Q in the range 1--5. From repeat measurements and comparisons
with other redshift catalogues, we find that redshifts with Q=3 have a
blunder rate (fraction of incorrect identifications) of just under 10\%
and Q=4 and~5 redshifts have blunder rates less than 1\%---the overall
blunder rate for reliable (Q$\ge$3) redshifts is 3\%. The overall rms
uncertainty in the Q$\ge$3 redshifts is 85\kms. The completeness of the
survey is computed as a function of both field and apparent magnitude.
The overall redshift completeness is 91.8\%, but this varies with
magnitude from 99\% for the brightest galaxies to 90\% for objects at
the survey limit.

The survey database has two components: a collection of FITS files, one
per object, which contain all the parameters and spectra for each
object, and a mSQL parameter database which can be used for
sophisticated searching, matching and sorting of the survey data. The
2dFGRS database is available through the survey WWW site at
http://www.mso.anu.edu.au/2dFGRS.
 
\section*{Acknowledgements}

The 2dF Galaxy Redshift Survey was made possible through the dedicated
efforts of the staff of the Anglo-Australian Observatory, both in
creating the 2dF instrument and in supporting the survey observations.

\end{document}